\documentclass[conference]{IEEEtran}
\usepackage{cite}
\usepackage{amsmath,amssymb,amsfonts}
\usepackage{algorithmic}
\usepackage{graphicx}
\usepackage{textcomp}
\usepackage[dvipsnames]{xcolor}
\usepackage[hyphens]{url}
\usepackage{comment}
\usepackage{bm}

\newcommand{\textoverline}[1]{$\overline{\mbox{#1}}$}

\usepackage{caption}
\usepackage[labelformat=simple]{subcaption}

\def\BibTeX{{\rm B\kern-.05em{\sc i\kern-.025em b}\kern-.08em
    T\kern-.1667em\lower.7ex\hbox{E}\kern-.125emX}}

\title{PIRM: Processing In Racetrack Memories } 
\author{Sebastien Ollivier, Stephen Longofono, Prayash Dutta, Jingtong Hu, Sanjukta Bhanja, Alex K. Jones}

\usepackage{multicol}
\usepackage{multirow}
\usepackage[ruled]{algorithm2e}
\SetAlCapSkip{0pt}
\usepackage{etoolbox}
\usepackage{xcolor}
\newbool{inccomment}
\setbool{inccomment}{true}
\newcommand{\sms}[1]{\ifbool{inccomment}{{\color{magenta}SMS: #1}}{}}
\newcommand{\akj}[1]{\ifbool{inccomment}{{\color{blue}AKJ: #1}}{}}
\newcommand{\sjl}[1]{\ifbool{inccomment}{{\color{red}SJL: #1}}{}}
\newcommand{\seb}[1]{\ifbool{inccomment}{{\color{ForestGreen}seb: #1}}{}}

\begin{document}
\maketitle
\thispagestyle{plain}
\pagestyle{plain}

\begin{abstract}

The growth in data needs of modern applications has created significant challenges for modern systems leading a ``memory wall.''  Spintronic Domain-Wall Memory (DWM), related to Spin-Transfer Torque Memory (STT-MRAM), provides near-SRAM read/write performance, energy savings and non-volatility, potential for extremely high storage density, and does not have significant endurance limitations.  However, DWM's benefits cannot address data access latency and throughput limitations of memory bus bandwidth.  
We propose PIRM, a DWM-based in-memory computing solution that leverages the properties of DWM nanowires and allows them to serve as polymorphic gates.  While normally DWM is accessed by applying spin polarized currents orthogonal to the nanowire at access points to read individual bits, \textit{transverse} access along the DWM nanowire allows the differentiation of the aggregate resistance of multiple bits in the nanowire, akin to a multi-level cell.  PIRM leverages this transverse reading to directly provide bulk-bitwise logic of multiple adjacent operands in the nanowire, simultaneously.  Based on this in-memory logic, PIRM provides a technique to conduct multi-operand addition and two operand multiplication using transverse access.  PIRM provides a 1.6$\times$ speedup compared to the leading DRAM PIM technique for query applications that leverage bulk bitwise operations.  Compared to the leading PIM technique for DWM, PIRM improves performance by 6.9$\times$, 2.3$\times$ and energy by 5.5$\times$, 3.4$\times$ for 8-bit addition and multiplication, respectively.  For arithmetic heavy benchmarks, PIRM reduces access latency by 2.1$\times$, while decreasing energy consumption by 25.2$\times$ for a reasonable 10\% area overhead versus non-PIM DWM.



\end{abstract}


\section{Introduction}
\label{sec:intro}

The growth in data needs of modern applications in the ``big data'' era has created significant challenges for modern systems.  While considerable effort has been undertaken to improve memory storage density and energy consumption---through deeply scaled memories and tiered memories that include non-volatile memory---the fundamental data access latency and throughput have not kept pace with application needs. 
This is commonly referred to as the ``memory wall''~\cite{villa2014scaling, mckee2004reflections}, as it limits potential performance of memory bound applications due to the limited bandwidth of the bus between memory and processor. Additionally, moving data on this bus has been proven to consume a disproportionately large amount of energy, especially for programs which require large working sets.  
For example, adding two 32-bit words in the Intel Xeon X5670 consumes 11$\times$ less energy than transferring a single byte from the memory to the processor~\cite{molka2010characterizing}.

Processing-in-memory (PIM)~\cite{elp2im,seshadri2017ambit,li2016pinatubo,yu2014energy,CNN_DWM} and near data processing (NDP)~\cite{li2017drisa, subramaniyan2017parallel} solutions promise to reduce the demands on the memory bus and can be a solution to efficiently realizing the benefit of increasingly dense memory from deep scaling and tiered memory solutions.  However, leading solutions for bulk-bitwise PIM in DRAM~\cite{elp2im, seshadri2017ambit} are limited to two operand operations.  Multi-operand bulk-bitwise PIM~\cite{li2016pinatubo} 
has been proposed for Phase Change Memory (PCM), a leading commercial candidate in the tiered memory space. However, its endurance challenges (circa 10$^8$ writes~\cite{wong2010phase}) and relatively high write energy (up to 29.7pJ per bit~\cite{luo2018write}) raise concerns about its effectiveness when used for PIM.  It is also popular to use the analog characteristics of, particularly memristor-based, crossbar arrays to accelerate neural networks~\cite{song2018graphr,yao2020fully} but these techniques can also lead to endurance as well as fidelity concerns.  Spin-transfer torque magnetic random-access memory (STT-MRAM) has been proposed for caches due to its SRAM-class access latency, non-volatility, favorable energy, and no appreciable endurance limitation.  STT-MRAM has also been proposed for PIM~\cite{kang2017memory}.  Unfortunately, STT-MRAM suffers from insufficient density to be competitive for main memories.

\textit{Spintronic Domain-Wall Memory} (DWM), often referred to as Racetrack Memory~\cite{Parkin-08-Science}, leverages the positive proprieties of STT-MRAM while dramatically improving the density by converting individual bits into multi-bit nanowires making it an appropriate choice for both main-memory integration and PIM.  
Two existing techniques propose using DWM for PIM with a goal of accelerating Convolutional Neural Network (CNN) applications. 
The first, abbreviated DW-NN for Domain-Wall Neural Network, leverages ``giant magnetoresistance'' to join a domain of  adjacent nanowires through a shared fixed layer.  This creates a shared access port that can be used to implement two operand \texttt{XOR} logic.  They also include a special sensing circuit between three nanowires used to compute a carry in order to create a ripple carry adder~\cite{yu2014energy}. Another, combines DWM with skyrmion logic to implement PIM, particularly addition and multiplication~\cite{CNN_DWM}, which we refer to as SPIM. 

Our paper proposes PIRM, or \textit{Processing In Racetrack Memories}.  PIRM achieves \textit{multi-operand} bulk-bitwise and addition PIM operations that can far outperform current DWM architectures for DRAM, PCM, and DWM.
Our approach treats the DWM nanowire as a \textbf{polymorphic gate} using a \textit{transverse} read~\cite{roxy2020novel} (TR), a method to \textit{access} the \textit{nanowire} and count the number of ones across multiple domains.  Using this ones count we can directly implement \textit{Racetracks} that are \textit{optimized} for \textit{computing} arbitrary logic functions, sum, and carry logic output.  
Presuming a sensing circuit that can distinguish data among seven domains, PIRM is able to directly compute seven operand bulk-bitwise operations and five-operand addition.  Multi-operand addition is particularly important to efficiently implement multiplication.  PIRM uses this building block to create \textit{specialized} PIM-enabled domain-block \textit{clusters} (DBCs).  These DBCs are interleaved throughout memory tiles to create the facility for massively parallel PIM in the PIRM main memory.  We show that the combined speedup of our multiplication procedure and the ability to process multiple operands significantly mitigates the memory wall and enables more sophisticated general PIM than prior work.
Specifically, the contributions of this paper are: 
\begin{itemize}
    \item We present a novel technique to utilize a segment of DWM nanowire as a polymorphic gate, including the required sensing and logic circuitry.
    \item We present the first technique, to our knowledge, to perform multi-operand logic and addition operations in DWM using this polymorphic gate.  We further extend this with shifting to implement efficient multiplication.
    \item We describe a PIM-enabled domain-block cluster architecture built from the PIM-enabled DWM.
    \item We propose a technique called Transverse Write (TW) to write and shift a segment of the nanowire in a single operation.
    \item We provide a detailed analysis of PIRM compared to state-of-the-art approaches in terms of energy, performance, and area. 
\end{itemize}

PIRM is effective for myriad applications such as database searching that requires multi-operand bulk bitwise computation and convolution-based machine learning that leverages arithmetic operations.

The remainder of this paper is organized as follows. In Section~\ref{sec:preliminaries}, we present the necessary background on Racetrack memory, its architecture, previous PIM techniques and TR. Next, Section~\ref{Sec:DesignOfPIRM} presents the basic concepts of PIRM, alongside our modified SA and supporting circuitry.   Furthermore, this section presents several approaches to perform smart multiplication with a concrete examples.  In Section~\ref{sec:experimentalResult}, we present the setup of our experiments to compare with state of the art DWM PIM architecture, including basic DWM architectures for addition and multiplication workloads and bulk-bitwise operations with Ambit and ELP$^2$IM.
Finally, we present our conclusions in Section~\ref{sec:conclusion}.

\section{Background and related work}
\label{sec:preliminaries}
In this section, we first introduce the fundamentals of DWM, how it functions, and our assumed memory architecture based on prior work.  We then discuss previous PIM work for both DWM and other memories. Finally, we describe details of TR and how it enables our PIM approach.

\subsection{Domain-wall Memory Fundamentals}
\label{sub_sec:DomainwallMemory}

\begin{figure}[htbp]
\centering
\includegraphics[width=\columnwidth]{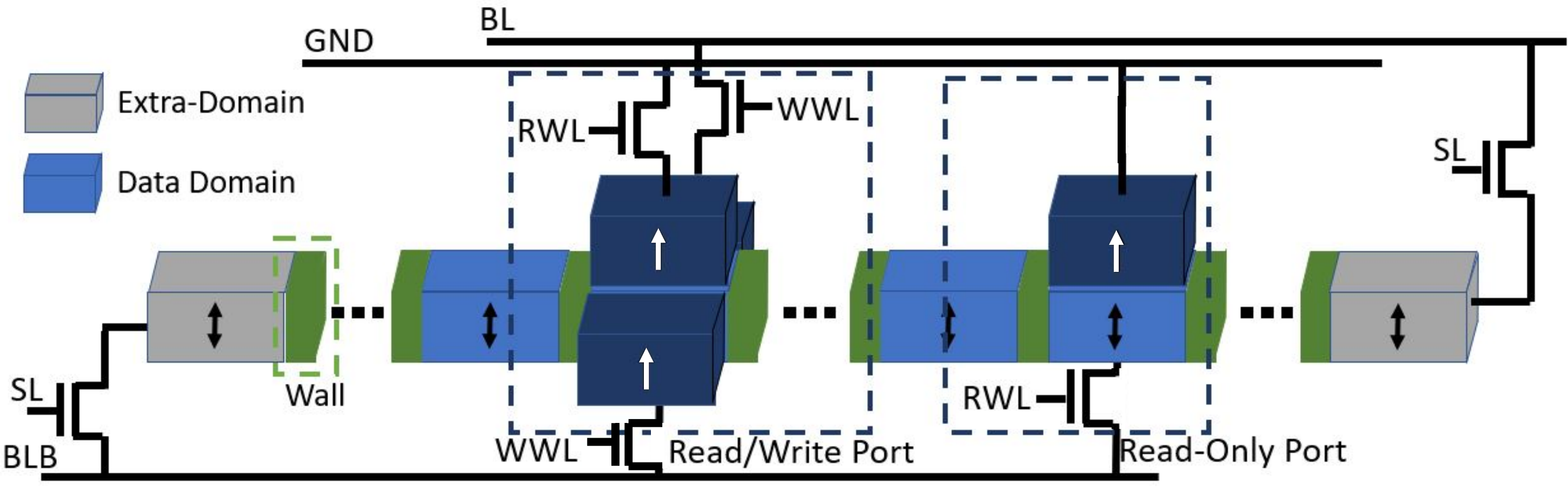}
\caption{Anatomy of a DWM nanowire. }
\label{DWMzoom}
\label{fig:nanowire}
\end{figure}

DWM is a spintronic non-volatile memory made of ferromagnetic nanowires. DWM nanowires consist of many magnetic domains separated by domain walls (DWs) as shown in Fig.~\ref{DWMzoom}. Each domain has its own magnetization direction based on either in-plane (+X/-X) or perpendicular (+Z/-Z) magentic anisotropy (IMA/PMA).  Binary values are represented by the magnetization direction of each domain, either parallel/antiparallel to a fixed reference. For a planar nanowire, several domains share one/few access point(s) for read and write operations~\cite{zhang2012perpendicular}. DW motion is controlled by applying a short current pulse laterally along the nanowire governed by \texttt{SL}. Since storage elements and access devices do not have a one-to-one correspondence, a random access requires two steps to complete:
\textcircled{1} {\em shift} the target domain and {\em align} it to an access port and \textcircled{2}
apply an appropriate voltage/current like in STT-MRAM to {\em read} or {\em write} the target bit.

The blue domains are dedicated for the actual data stored in memory. The grey domains are extra-domains used to prevent data loss while shifting data into the nanowire towards the extremities. The dark blue elements are the read or read/write ports.  Fig.~\ref{DWMzoom} contains a read-only port that has a fixed magnetic layer, indicated in dark blue, which can be read using \texttt{RWL}.  The read/write port is shown using shift-based writing~\cite{DWM_Tapestri} where \texttt{WWL} is opened and the direction of current flows between \texttt{BL} and \textoverline{\texttt{BL}}, and reading conducted from \textoverline{\texttt{BL}} through the fin and up through \texttt{RWL} to \texttt{GND}. 

To align a domain with an access port, a current needs to be sent from one extremity of the nanowire, shifting each domain. This inherent behavior of DWM can be imprecise, generating what is known as a ``shifting fault'' in the literature. Several solutions have been proposed to mitigate this fault mode~\cite{ollivier2019dsn, hifi, archer2020foosball}.

\subsection{Memory Designs with DWM}
\label{sec:DWM-memory-arch}

\begin{figure*}[t]
     \centering
     \begin{subfigure}[b]{0.42\linewidth}
     \centering
     \includegraphics[width=\columnwidth]{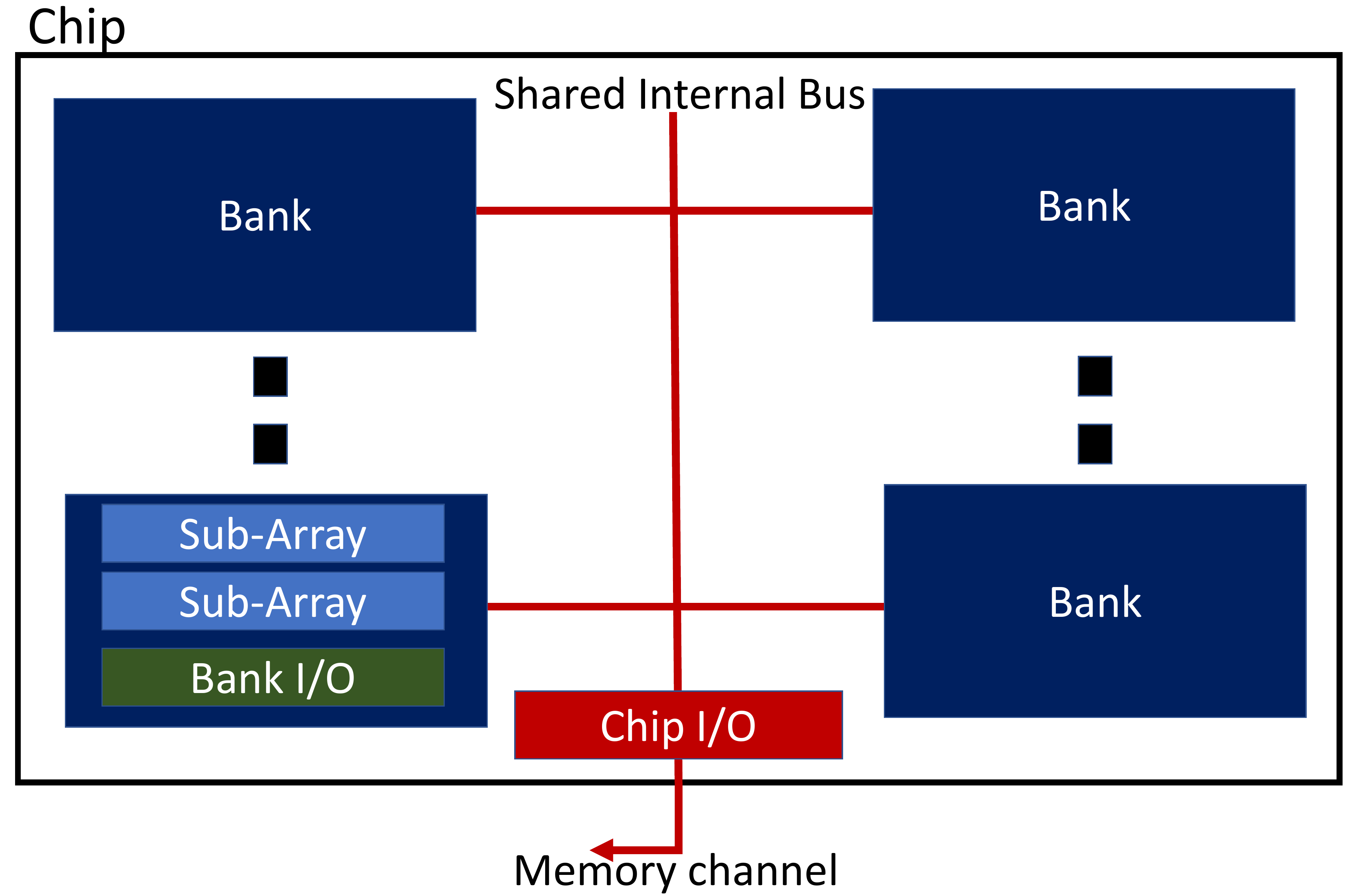}
     \caption{Chip concept retained from DRAM.}
     \label{fig:mem-top-level}
     \end{subfigure}
     \begin{subfigure}[b]{0.42\linewidth}
     \centering
         \includegraphics[width=\columnwidth]{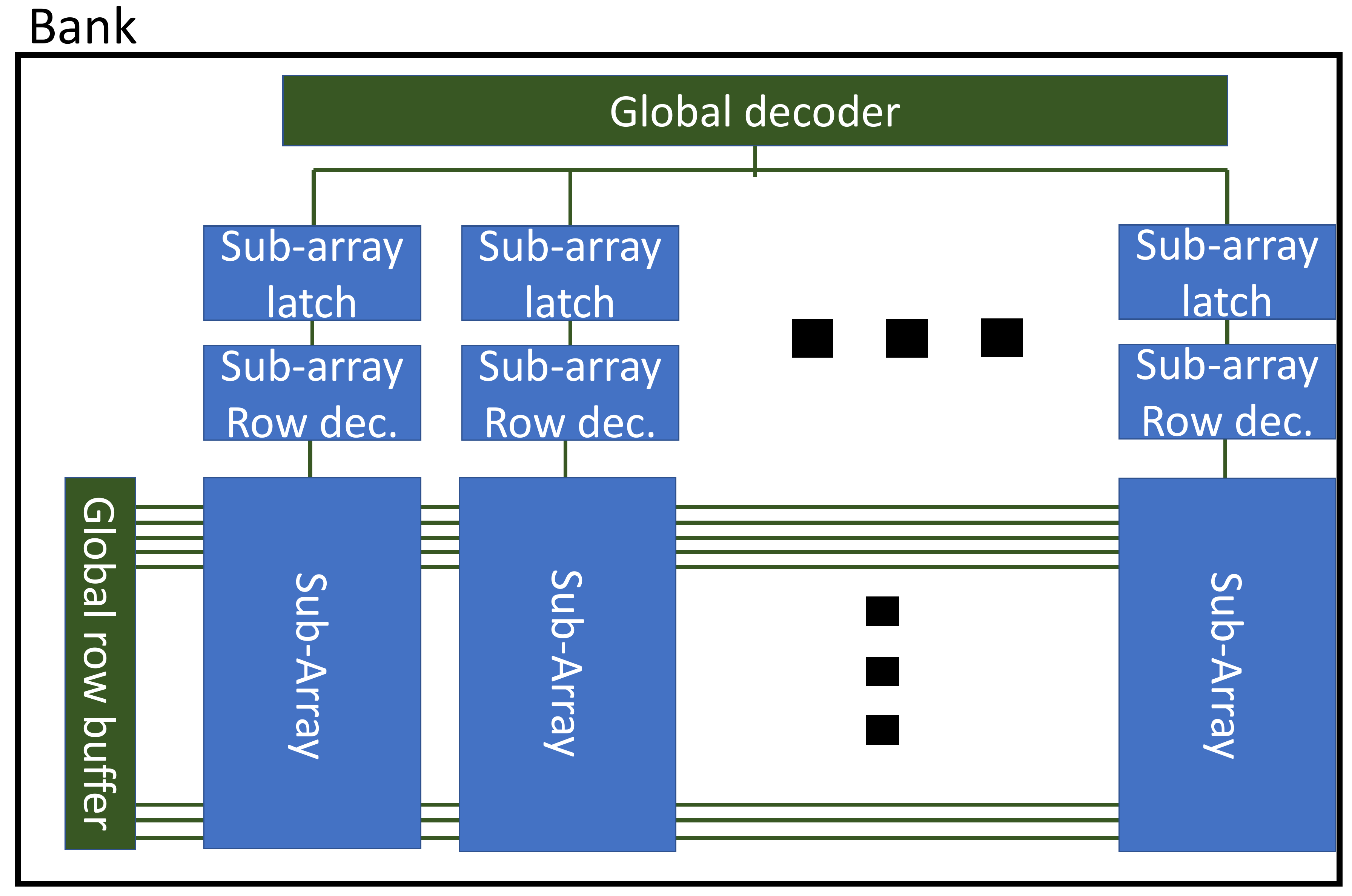}
         \caption{DWM Bank}
         \label{fig:Bank}
     \end{subfigure}
     \begin{subfigure}[b]{0.42\linewidth}
         \centering
         \includegraphics[width=\columnwidth]{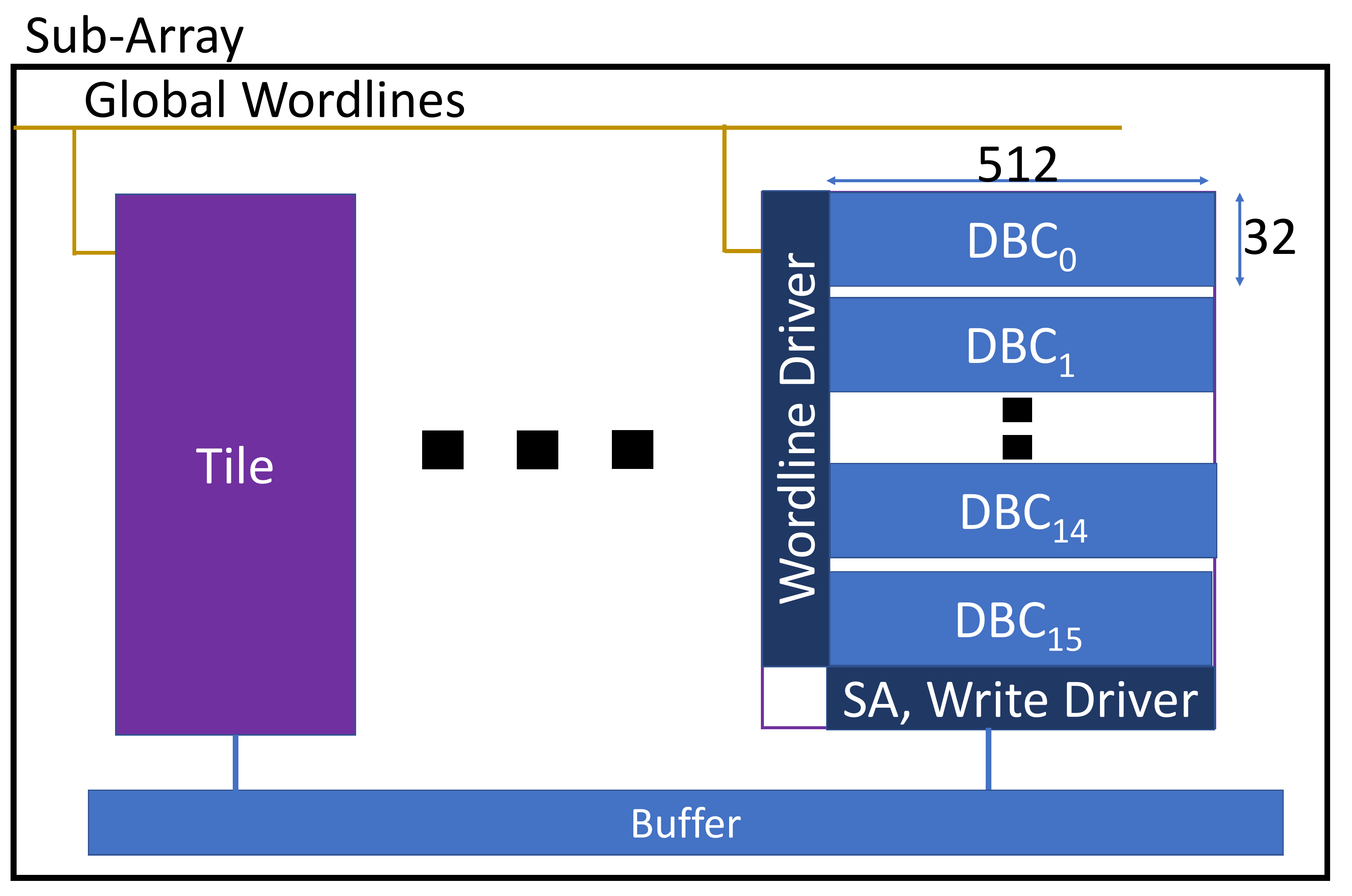}
         \caption{DWM subarray}
         \label{fig:SubArray}
     \end{subfigure}
     \begin{subfigure}[b]{0.42\linewidth}
     \centering
         \includegraphics[width=\columnwidth]{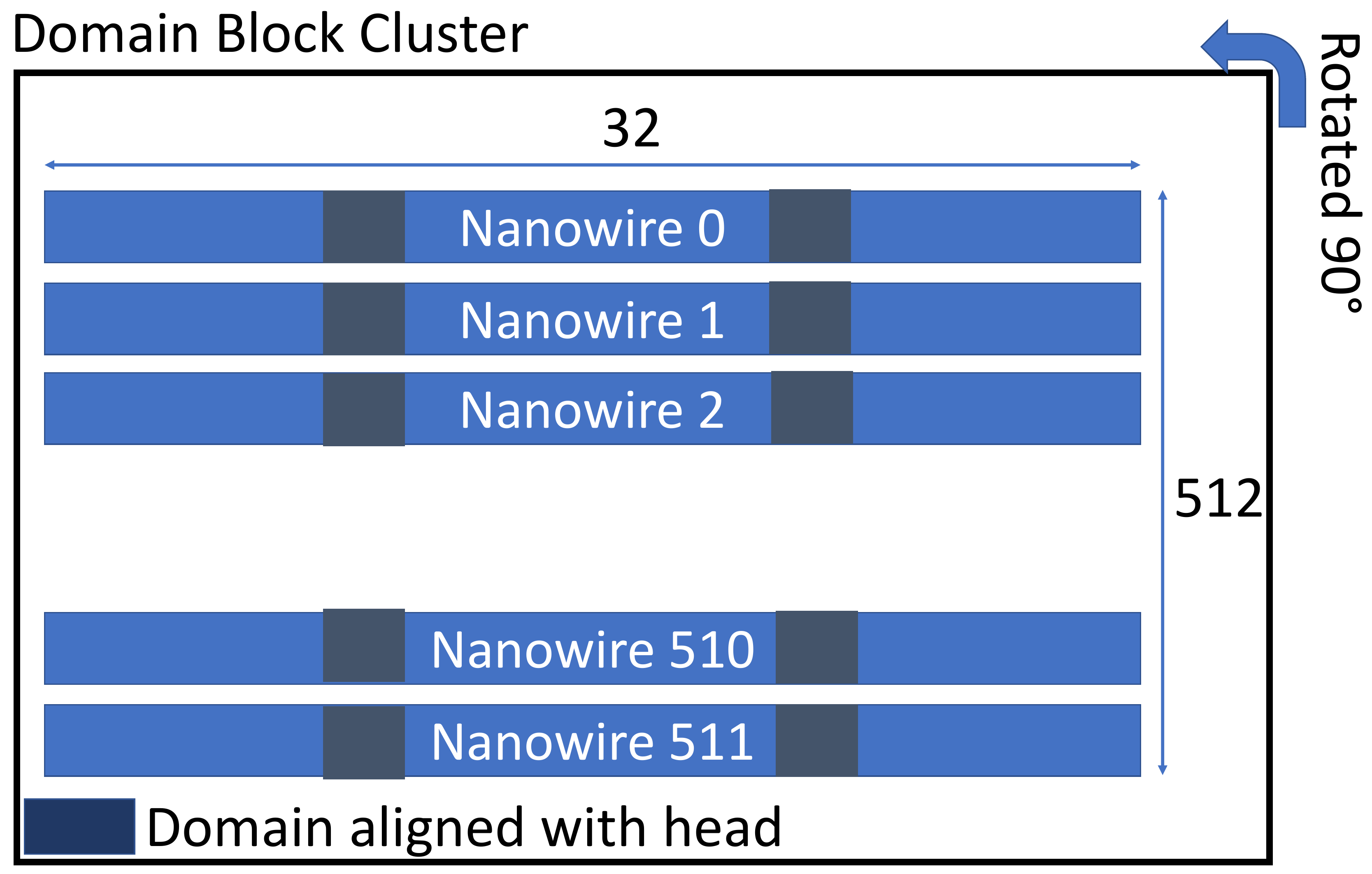}
         \caption{DWM Domain Block Cluster (rotated 90$^\circ$)}
         \label{fig:DBC}
     \end{subfigure}
        \caption{DWM Architecture}
        \label{fig:three graphs}

\end{figure*}

To employ DWM as main memory, we propose to use a similar architecture that has been previously proposed in the literature for emerging technologies (including DWM).  This architecture maintains the same I/O interface as a traditional DRAM memory hierarchy to simplify adoption of a new technology.  Thus, this architecture preserves the bank and subarray organization as shown in Fig.~\ref{fig:mem-top-level} and the tile size~\cite{kim2012case} as shown in Fig.~\ref{fig:Bank}.   This also allows for data movement within the memory such as inter-bank copying that is described in previous work~\cite{seshadri2013rowclone}.
Fig.~\ref{fig:SubArray} shows a DWM subarray broken down into tiles, which share the same global wordlines.  To facilitate DWM integration we divide tiles into domain block clusters. Each DBC shares the same local sensing circuitry and write driver as described for other main memory architectures using non-volatile memory~\cite{yue2013exploiting}.

Each DBC, expanded and turned 90$^\circ$ so that the nanowires are shown horizontally in Fig.~\ref{fig:DBC}, consists of $X$ parallel racetracks composed of $Y$ data domains.  Thus, $X$ represents the number of bits that can be accessed simultaneously.  We show a typical example where $X=512$.  $Y$ represents the distinct row addresses contained within the DBC.  $Y$ is limited as the necessary current required for successful shifting increases with the number of domains. 
Moreover, larger values of $Y$ introduce increasingly large access delay due to the longer average travel distance.
We show a typical example of $Y=32$.  $Y$ is unlikely to exceed 64 or 128 as reported in the literature~\cite{TapeCache, sun2013cross}.  This architecture allows $32 \leq Y \leq 512$, which allows sufficient scaling for longer nanowires.  

While, a single access point is necessary for each nanowire in the DBC, adding heads can reduce the delay by reducing the shift distance between accesses~\cite{TapeCache}.  We show two heads in the example, which would traditionally divide the nanowire length into equal sections to minimize shift latency and reduce the number of overhead domains required.  In addition, a second access port can also enable the polymorphic gate capability of the nanowire if placed sufficiently close to allow transverse access.  We discuss this further in Section~\ref{sub_sec:TransverseRead}.  However, first we review prior work in PIM in the next section.


\subsection{Processing in Memory}
\label{sec:PIM-Background}
In this sub-section, we first present the state of the art bulk-bitwise operations in DRAM, followed by related works proposed for PIM in DWM.

\subsubsection{PIM with bulk-bitwise operations in DRAM and PCM}
\label{sub_sub_sec:BulkBitwise}
There have been two major proposals to conduct bulk-bitwise processing directly in DRAM.  Bulk-bitwise logic combines two rows ``bitwise'' with the same logic operation such that $c_i = a_i$ \texttt{OP} $b_i$.  Ambit proposed to open three DRAM rows simultaneously and compare the combined voltage to the sensing threshold, \textit{i.e.,} $\frac{V_{DD}}{2}$~\cite{seshadri2017ambit}.  A majority of `1's results in $\geq \frac{2 V_{DD}}{3}$ which would drive the senseamp (SA) to $V_{DD}$.  A minority of `1's results in $\leq \frac{V_{DD}}{3}$ driving the SA to $V_{CC}$.  Thus, to compute an \texttt{AND} the third row is used as a control row set to `0' so that both data rows must contain `1's for the result to be `1.'  \texttt{OR} is computed setting the control row to `1' requiring only one data row to be `1.'  This process is destructive, as all three rows now contain the result of the logical operation.  

Ambit uses a technique called RowClone~\cite{seshadri2013rowclone} which essentially opens the source row, waits for the SA to refresh the row and then opens the destination row which is overridden by the SA just as happens in a write.  Thus, operands are duplicated in a safe location to conduct the logic operation without destroying the orginal data.  To create a complete logic set a dual-contact cell (DCC) concept is employed allowing a cell to be read as the inverted value through \texttt{$\overline{\text{BL}}$}.  A DCC row requires the same overhead as two regular rows.  To execute $A$ \texttt{XOR} $B$ they leverage these DCC rows to first compute $k = A$ \texttt{AND} $\overline{B}$, followed by $k' = \overline{A}$ \texttt{AND} $B$, and finally $k$ \texttt{OR} $k'$

ELP$^2$IM improves on Ambit by directly performing logic operations without moving the data.  Instead, the technique changes the pseudo-precharge state of the SA to replace the control row~\cite{elp2im}.  The process requires multiple comparisons to ultimately determine the final logic value, but avoids the need for cloning rows.   They demonstrate a 3.2$\times$ performance improvement over Ambit as well as recent near-data processing approaches~\cite{li2017drisa} for bitmap and table scan applications. 

Pinatubo is a PIM technique for PCM that resembles aspects of Ambit and ELP$^2$IM.  Like Ambit, it opens the two rows for comparison simultaneously, and like ELP$^2$IM it adjusts the sensing circuitry to conduct different operations~\cite{li2016pinatubo}.  For example, changing the $V_{TH}$ to $<\frac{V_{DD}}{2}$ allows an \texttt{OR} operation and $>\frac{V_{DD}}{2}$ allows an \texttt{AND} operation.

Ambit, ELP$^2$IM, and Pinatubo insomuch as they are complete logic sets are capable of computing more complex arithmetic operations.  More complex logic requires sequential steps to determine the result similar to the \texttt{XOR} example described for Ambit.  Next we discuss PIM for DWM.

\subsubsection{PIM in DWM}
\label{Sub_sub_sec:PIMinDWM}

DW-NN creates a PIM processing element with dedicated circuitry to support current passing through two stacked domains at once. This allows measurement of the aggregate giant magnetoresistance (GMR) across the stacked domains~\cite{yu2014energy}.  This computes \texttt{XOR} which is `0' if the data is parallel and `1' if the data is anti-parallel.  To conduct addition, operand bits are stored in consecutive bits within a single nanowire.  The \texttt{XOR} operation is used in combination with a pre-charge sensing amplifier (PCSA) that can compute a function of data from three nanowires' access port---sum $S$ is the result of two consecutive \texttt{XOR}s, and $C_{OUT}$ is the result of the comparison of $\text{PCSA}(A,B,C_{IN}) > \text{PCSA}(\overline{A},\overline{B},\overline{C_{IN}})$.  Both operations are bitwise serial since they must be shifted into alignment with the GMR/MTJs.
Because operands are stored within a single nanowire, multiplication is possible using addition of shifted versions of one operand.  Compared to accessing data and using a general purpose processor for computation, DW-NN claims an energy improvement of 92$\times$ and a throughput improvement of 11.6$\times$ for an image processing application.

SPIM extends DWM storage with dedicated skyrmion-based computing units~\cite{CNN_DWM}. Within these units, custom ferromagnetic domains are physically linked together with channels that support \texttt{OR} and \texttt{AND} operations.  By permanently merging many such domains and channels, full adder circuits are formed to perform addition and multiplication.  SPIM demonstrates 8-bit addition and multiplication and reports being two times faster, while decreasing energy and area by 42\% and 17\%, respectively compared to DW-NN.  Savings are achieved by parallel bulk-bitwise ops not possible in DW-NN and reduced area compared to a similar ALU implemented with CMOS.



DW-NN is capable of \texttt{XOR} addition, and multiplication, but is bit-serial and cannot compute other logic functions.  SPIM improves on DW-NN with skrymion logic units, but with more complex dedicated hardware and a more restrictive set of operations.  PIRM provides a new method to do PIM in DWM that can achieve a superset of all the instructions proposed among these various methods while increasing the parallelism and efficiency.  To accomplish this, PIRM leverages the transverse read operation discussed next.

\subsection{Transverse read}
\label{sub_sec:TransverseRead}

TR was recently proposed~\cite{roxy2020novel} and leveraged to improve reliability via detection and correction of over/under-shifting faults~\cite{ollivier2019dsn}. TR is akin to using a portion of a DWM nanowire as a multi-level STT-MRAM cell. Specifically, the idea is to read an aggregate function of several domains at once along the nanowire. The output of a TR provides the number of ones stored between two heads or one head and an extremity, but without information about their positions.  As the number of domains in the TR increases, the minimum sense margin decreases which creates a limit on the number of domains that can be included in a TR.  We refer to this as the \textit{maximum TR distance} or simply TRD.  

\begin{figure}[t]
\centering
\includegraphics[width =\columnwidth]{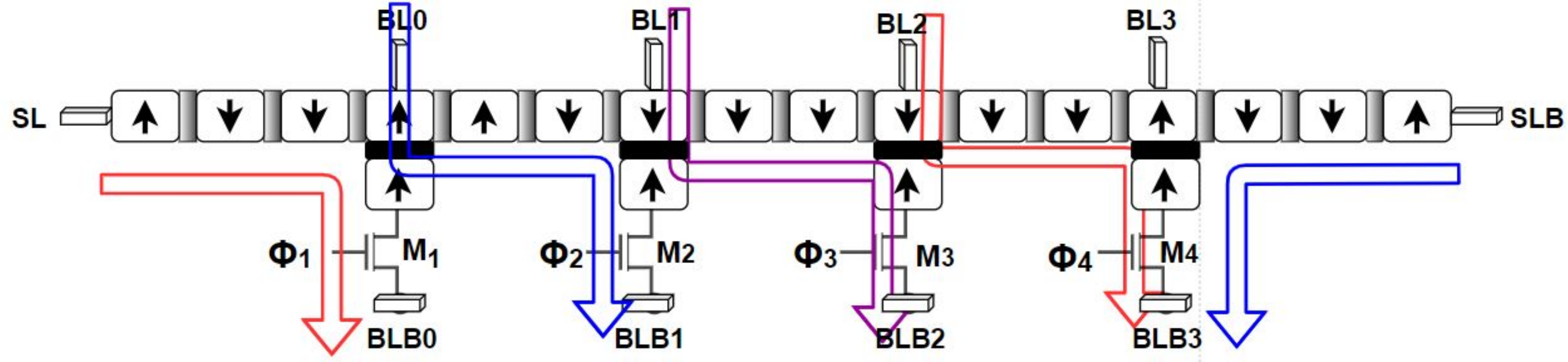}
\caption{Transverse read example for a full nanowire }
\label{TR}
\label{fig:TR}
\end{figure}
Fig.~\ref{TR} presents a segmented TR that can be used to query the full nanowire in the case that distance from the extremity to the access point is larger than the TRD.  Each colored arrow represents the path taken by the current used to perform the TR. For instance, to perform a TR on the middle four domains (purple arrow), the transistors \texttt{M1}, \texttt{M2} and \texttt{M4} are open and \texttt{M3} is closed, thus when a current is sent from \texttt{BL1}, it has to go though the four middle domains and exit through \texttt{M3}. 

The two red and blue arrows indicate TR over regions with the same color can occur simultaneously. For the red arrows, \texttt{M2}, \texttt{M3} and \texttt{SLB} transistors are open while \texttt{M1} and \texttt{M4} are closed. The current sent from \texttt{SL} and \texttt{BL2} will flow to \texttt{M1} and \texttt{M4}, respectively, reading two and one `1's as output, respectively. Due to the larger nanowire resistivity between \texttt{BL0} and \texttt{BL2}, the leakage current is small enough that the TR can safely be parallelized.

In the next section, we present how the TR can be used to perform logical and arithmetic operations, forming the foundation of PIRM.

\section{PIRM}
\label{Sec:DesignOfPIRM}

In this section, we discuss the PIRM architecture; this includes (i) modifications to the sensing circuit leveraging TR to realize multi-operand PIM, and (ii) the algorithms to achieve multi-operand addition and two-operand multiplication.

\subsection{PIRM Architecture}
\label{sub_sec:ArchitectureModifications}

We propose to add PIM capability to a portion of the DBCs in the memory architecture, see Fig.~\ref{fig:SubArray}.  The number of PIM enabled DBCs can be tuned based on the overhead and the desired PIM parallelism.  The detailed DBC shown in Fig.~\ref{fig:DBC} shows two access points.  In PIRM, these access points are spaced according to the TRD.  While TR has been demonstrated for a conservative $\text{TRD}=4$~\cite{roxy2020novel} our experiments using the LLG magnetic simulator~\cite{LLG} and the resulting resistance levels indicate that TR can be scaled to a $\text{TRD}=7$ by increasing the sensing current.  This yields eight resistance levels which can be encoded by three bits.  We will see how $\text{TRD}=7$ is particularly effective for addition.  $\text{TRD}=7$, while optimistic, is significantly more realistic than the presumed $\text{TRD}=32$ assumed in prior work~\cite{ollivier2019dsn}.  

Assuming the  distinct  row  addresses  contained within  the  DBC $Y=32$ as discussed in Section~\ref{sec:DWM-memory-arch}, and  with a single access point, this nanowire would require 63 domains ($2Y-1$) to permit shifting data at the extremities to the access point.  Normally, adding a second access point would place ports at positions 9 and 25, reducing the number of overhead domains from 31 to 16.  To enable TR with a $\text{TRD}=7$, the ports would move to positions 14 and 20 and the overhead domains would only reduce from 31 to 25.  Adding ports in this way provides some reduction of average shift distance while allowing for the TR operation.  For two ports to remain at their optimal shift reduction position would not realistically allow a TR between them, because their distance would be 14 domains.  Our experiments did not support a such a $\text{TRD}=14$ at this time to be feasible.

\begin{figure*}[t]
\centering
\begin{subfigure}[b]{3in}
\centering
    \includegraphics[width=\columnwidth,angle=90,origin=c]{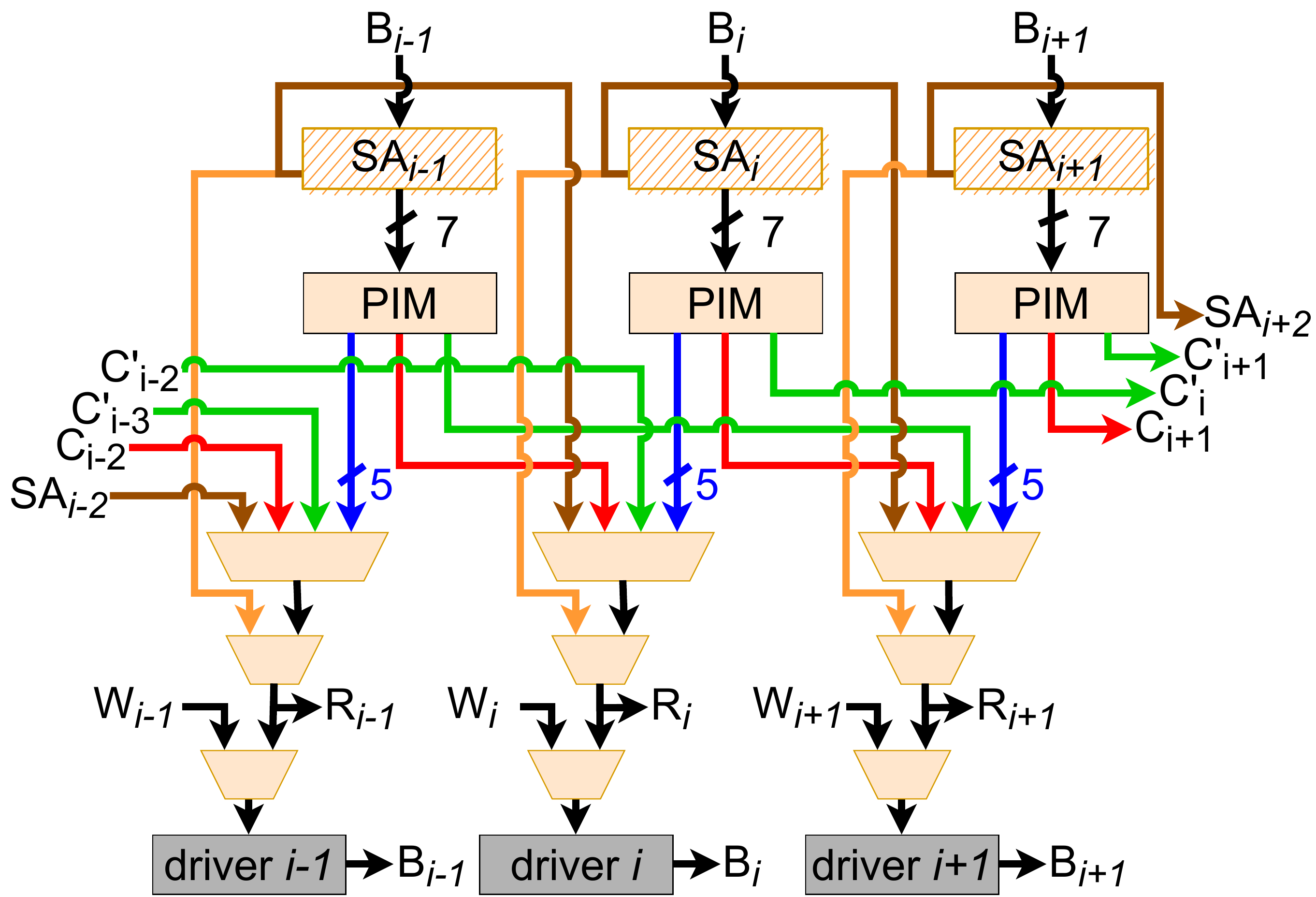}
    \caption{PIM DBC SA and driver circuit.}
    \label{fig:subarray_micro_arch}
\end{subfigure} 
\begin{subfigure}[b]{3in}
\centering
\includegraphics[width=\columnwidth]{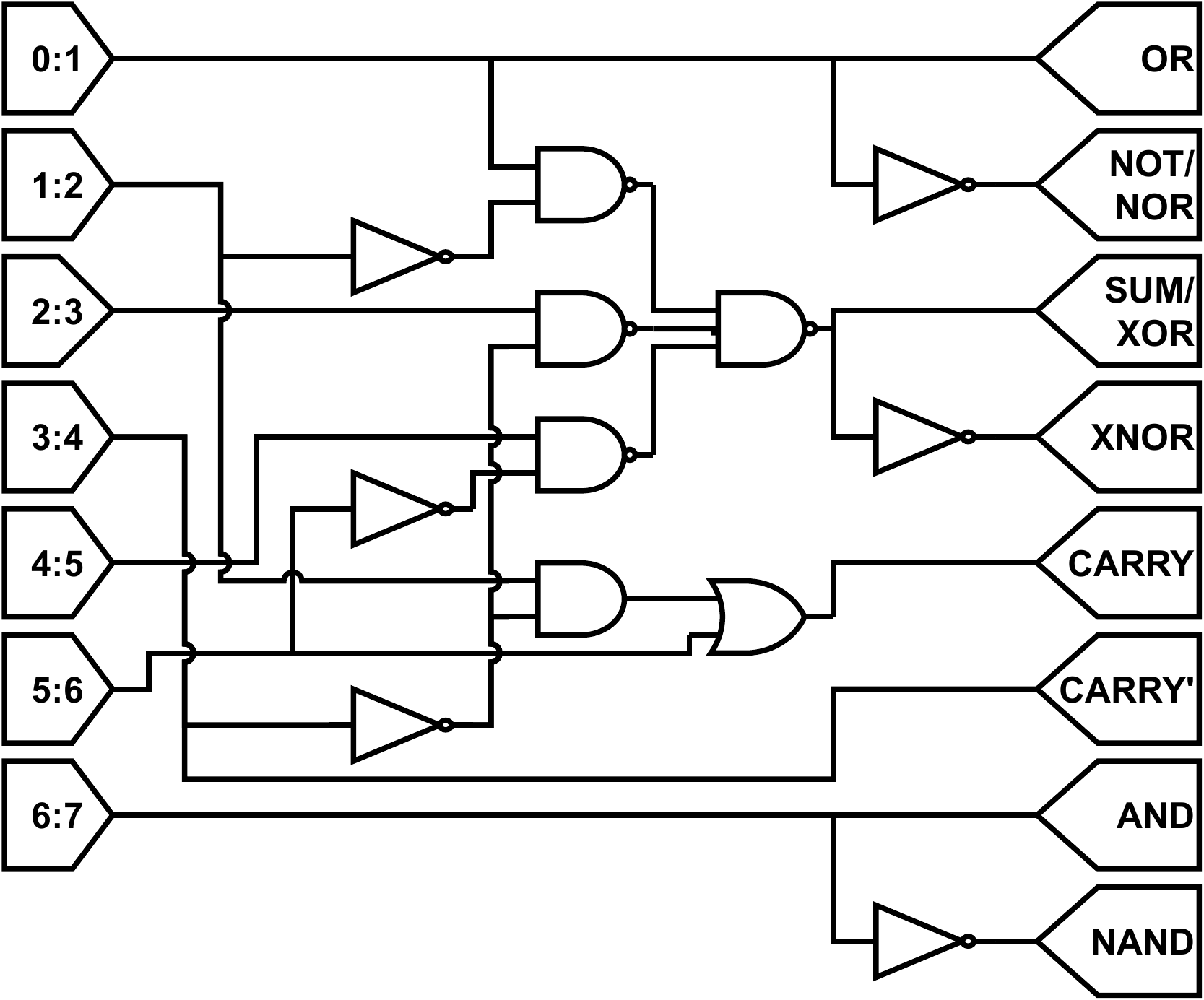}
\caption{PIM block logic.}
\label{fig:PIMgates}
\end{subfigure} 

\caption{Overview of the PIM enabled DBC extensions.}
\end{figure*}

\begin{figure}[tbp]
\centering
\includegraphics[width=.9\columnwidth]{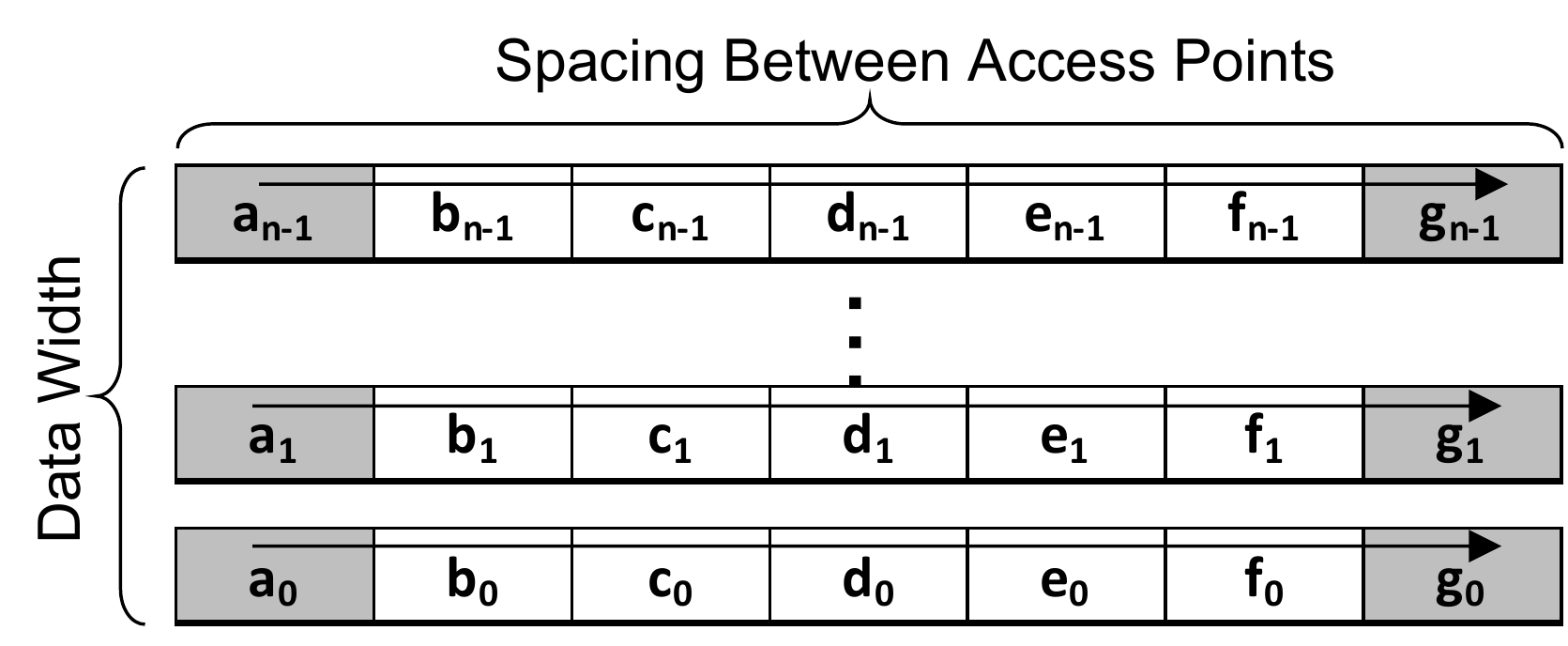}
\caption{TR for Bulk-Bitwise Operations }
\label{fig:BulkBitwiseOperations}
\end{figure}

For a tile with the additional access port to conduct TR, we modify the sensing circuitry as shown in Fig.~\ref{fig:subarray_micro_arch} where the tan blocks show the added elements.  To enable performing TR requires each SA$_i$ to output seven level bit values such that SA$_i$[$j$] is `1' if there are $\geq j$ `1's in the TR and $j \in 1..7$.  The extension with additional sensing circuitry is represented by a hashed tan block.  These SA outputs become the seven inputs for the PIM unit described in Fig.~\ref{fig:PIMgates}. The PIM logic output is selected from a multiplexer.  Note, the $i$th multiplexer selects some values provided by the local PIM block and some that come from $(i-1)$st and $(i-2)$nd PIM blocks.  We will explain the purpose for the color coding and these connections in the following sections.  

There is a direct read from the SA shown in orange that bypasses the PIM logic and the selector to a single two-way mux that feeds the read port.  Thus, either a direct read or result of PIM logic can be directly forwarded via the hierarchical row-buffer structure to the memory controller and returned to the processor.  The capability is also added that PIM output can be written back to the memory block so an additional selector multiplexes the PIM logic with the write port input.  As in previous work~\cite{seshadri2017ambit,seshadri2013rowclone,li2016pinatubo}, given the hierarchical row buffer in the memory, the shared row buffer in the subarray or across subarrays can be used to move data from non-PIM DBCs to PIM-enabled DBCs.  

\subsection{PIRM Multi-operand Bulk-bitwise Operations}
\label{Sub_sec:PIMDWoperations}


\label{sub_sec:Between2and5}

The TR operation described in Section~\ref{sub_sec:TransverseRead} allows direct implementation of bulk-bitwise operations.  But to accomplish this requires the additional sensing capability discussed in the last section and a small amount of additional logic shown in Fig.~\ref{fig:PIMgates}.  If the TR level is above one, the \texttt{OR} operation is `1' and similarly the inversion of this reports \texttt{NOR}.  If a single data value is stored and the remaining rows are zero padded, this output also reports \texttt{NOT}.  \texttt{AND} and \texttt{NAND} are obtained in a similar fashion but with the highest TR level.  \texttt{XOR} reports exclusively the odd TR levels computed by relatively simple NAND/NAND implementation.  To save area, \texttt{XNOR} is the inverted value of \texttt{XOR}.  While, at first glance, this might appear like a significant overhead, its important to keep in mind that this logic will compute these operations for seven operands in parallel.  To support addition, the PIM block also contains a carry $C$ computation which is a function of TR levels above two and not above four or above six.  A super carry $C'$ is computed from TR level above 4 and sum $S$ is equivalent to \texttt{XOR}.  Details on the energy consumption and area overhead are discussed in Section~\ref{sec:experimentalResult}.

We show an example in Fig.~\ref{fig:BulkBitwiseOperations}, for the portion of the DBC between the two access points, denoted by shaded domains where $a$ or $g$ could be directly read, and the rest of the nanowire is abstracted away for convenience of display.  Using a TR, a multi-operand operation can be directly obtained for bulk-bitwise \texttt{OR}, \texttt{NOR}, \texttt{AND}, \texttt{NAND}, \texttt{XOR}, or \texttt{XNOR}.  Because \texttt{OR} uses the same sensing circuit as a traditional read, but the read is conducted using a TR, it is made available through the orange path; the remaining five operations are denoted by the blue output of the PIM block [Fig.~\ref{fig:subarray_micro_arch}].  Comparing fewer than seven operands can be accomplished by zero-filling the unused locations in the scope of the TR.  The result can be written over one of the original operands (either $a$ or $g$) or written into a separate DBC.
In order to minimize the energy and area overhead PIRM will apply this PIM extension to a subset of the tiles, generally one tile per subarray.
\subsection{PIRM Multi-operand Addition}
\label{sub_sec:addition}



\begin{figure}[tbp]
\centering
\includegraphics[width=.9\columnwidth]{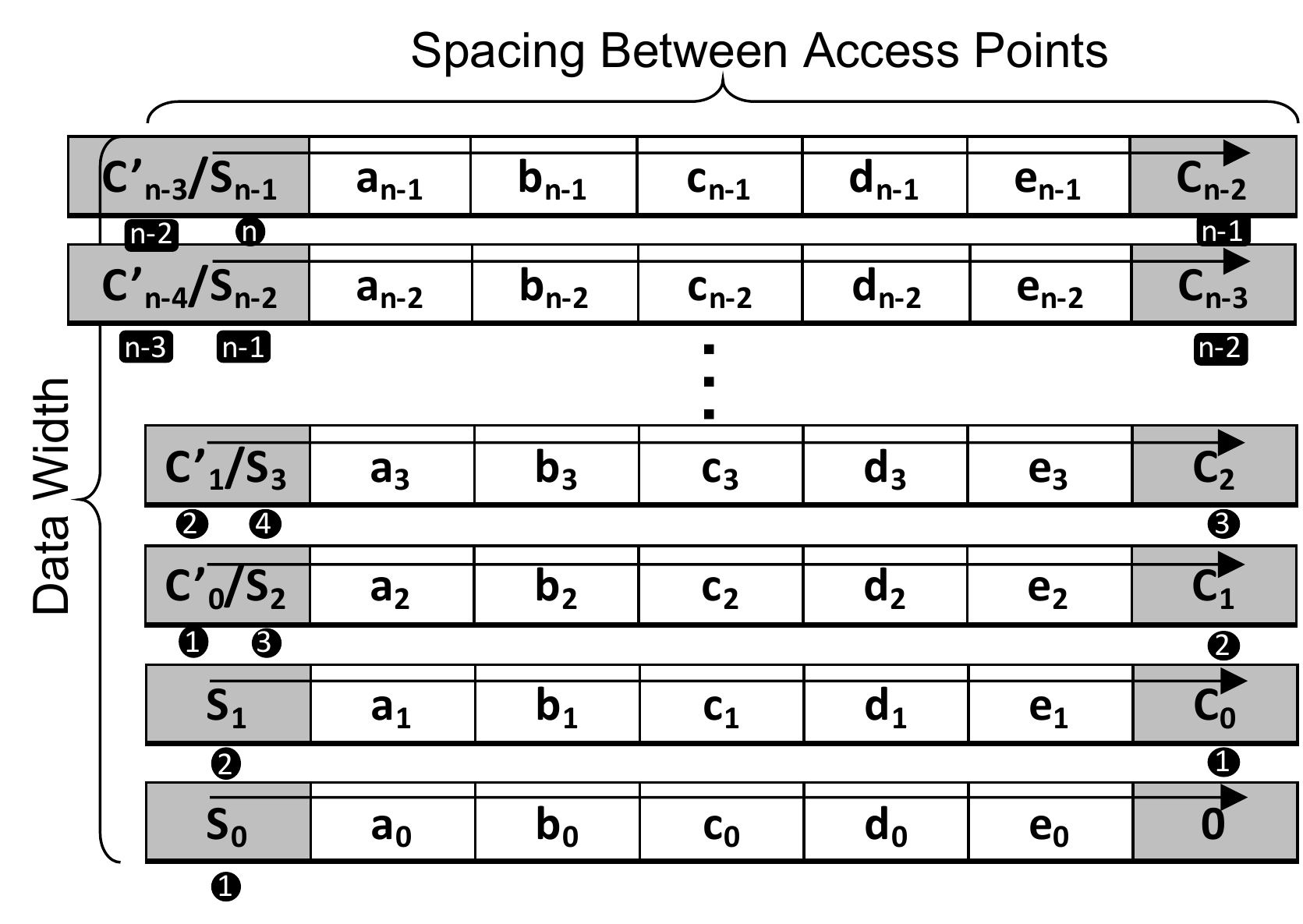}
\caption{Addition using TR}
\label{fig:add-simple}
\end{figure}

Based on the bulk-bitwise operations from the previous section, we show an example addition operation for five operands in Fig.~\ref{fig:add-simple}.  In step \textcircled{1}, referencing Fig.~\ref{fig:subarray_micro_arch}, a TR of $dwm_0$ (first nanowire) is conducted, evaluating $bit_0$ of all operands.  $S_0$, which is \texttt{XOR} of $a_0...e_0$, computed by the PIM block and is among the five blue bits.  
Simultaneously, carry, $C_0$, is computed and sent to the right to the driver for $dwm_1$ shown in red and super carry, $C'_0$ is sent to the driver for $dwm_2$, shown in green in Fig.~\ref{fig:subarray_micro_arch} and computed using the logic functions shown in Fig.~\ref{fig:PIMgates}.  $S_0,C_0,C'_0$ are written into the left access port ($port_L$) of $dwm_0$, the right access port ($port_R$) of $dwm_1$, and $port_L$ of $dwm_2$ simultaneously.  In step \textcircled{2}, a similar set of steps occurs except the operations include $C_0$ in addition to $a_1...e_1$.  Then in step \textcircled{3} TR is conducted over $C'_0,a_2...e_2,C_1$, seven total elements.  In the general case, for step $k+1$ (\textit{i.e.,} $dwm_k$), TR is conducted over $C'_{k-2},a_k...e_k,C_{k-1}$ with $S_k$ written to $port_L$ of $dwm_k$, $C_k$ written to $port_R$ of $dwm_{k+1}$ and $C'_k$ written to $port_L$ of $dwm_{k+2}$.  The control for this operation is a simple counter circuit that provides selectors values for a window of three nanowires and activates the bit lines for $k...k+2$.  

Because the carry chain requires keeping the $port_L$ and $port_R$ clear to write $C,C'$, for a $\text{TRD}=7$ we can compute a maximum of five-operand addition.    While it may be possible in the future to increase $\text{TRD}\geq8$, we explore a technique to efficiently add more than five operands in the context of PIRM multiplication, which we discuss in detail next.

\subsection{PIRM Multiplication}
\label{sub_sec:multiplication}

The capability to compute addition in memory, as described in the previous section, provides the foundation for multiple ways to conduct multiplication.  A particularly na{\"i}ve method to compute $A * B$ is to add $A$ $B$ times.  For example, $3A$ can be computed as $A+A+A$. Thus, we can perform a multiplication by doing several additions.  This method can quickly exceed the capacity of a single multi-operand add.  Consider $9A$, this can be computed by computing $5A$ in one step, and then computing $5A+A+A+A+A$.  While this method could be improved by generating $5A$ in one addition step, then replicating $5A$ to generate $25A$, and so on, this method is clearly inefficient.  One method to accelerate this process is to shift the copies of $A$ to more quickly achieve the precise partial products that, when summed, produce the desired product.  

In Fig.~\ref{fig:subarray_micro_arch}, we show how data read from bit $i$ is forwarded to bit $i+1$ using the brown lines.  These lines originate from the same place as the orange lines from $i$ but are shown coming from the opposite side of the SA for reduced clutter in the figure.  This connection allows a logical left shift which is equivalent to a multiply by 2 or $A'$.  To logically shift by more than one position, we first write $A'$ and then shift and write $A''$ or $A<<2$.  It is important to distinguish between these \textbf{logical shifts} being discussed, which move bits \textit{between} nanowires and \textbf{DW shifts}, which shift the nanowires to access different data locations.  Looking at Fig.~\ref{fig:add-simple}, logical shifts occur in the Y direction and require the multiplexing logic from Fig.~\ref{fig:subarray_micro_arch}, DW shifts move in the X direction.  So, to write $A<<k$ requires $k$ shifted read (brown arrows) and write operations.  However to write $A<<k$ next to $A<<k+1$ requires $k$ shifted read and followed write steps, an additional shifted read, a DW shift, followed by a write.  Thus, to write $y$ shifted copies with a max logical distance of $x$ requires $x-1$ shifted read/write operations and $y$ DW shifts.   Based on the logical shifting capabilities we describe techniques to leverage PIRM to conduct efficient multiplication in different situations.

\subsubsection{Constant Multiplication}
When one of the operands of the multiplication operation is known, there are several ways to efficiently take advantage of PIRM to complete  multiplication leveraging shifting.  At compilation time, a method based on Booth multiplication is possible~\cite{lefevre-multiplication,Bernstein-multiplication} where numbers can be represented using \texttt{0}, \texttt{N}, and \texttt{P}, which represent 0, -1, and 1, respectively. For example, consider a constant multiplier 20061 $\rightarrow$ ``100111001011101,'' this can be encoded as \texttt{P0P00N0P0N00N0P}.  It can be decomposed using the pattern \texttt{P000000P0N}, which corresponds to 515, in positive and negative forms shifted by different amounts: \texttt{P000000P0N00000} $-$ \texttt{P000000P0N} $+$ \texttt{00P000000000000} $=$ \texttt{P0P00N0P0N00N0P}.  

Thus, 20061 times $A$ can be computed in two addition steps: \textcircled{1} $A << 9 + A<<1 + A \rightarrow 512 A + 2 A + A = 515 A$, \textcircled{2} $515 A << 5 - 515 A + A<<12 \rightarrow 16480 A - 515 A + 4096 A = 20061 A$.  Note $-515 A$ can be computed by generating $\overline{515 A} + 1$ making the last step $515 A << 5 + \overline{515 A} +1 + A << 12$ which is still one addition step.  This is a significant improvement over adding 20061 copies of $A$.  

\subsubsection{Arbitrary Multiplication} A more generic method that can also work for arbitrary multiplications is to use the `1's in the multiplier to denote which shifted copies of the multiplier to be summed to create the product.  In the 20061 example, there are 9 `1's in the binary form of 20061.  Thus, the method to directly compute the product is logically shift $A$ $n$ times where $n$ is the bit-width of the multiplier $B$.  When $b_i =$ `1' then we also shift the nanowire to retain that ``partial product.''  When five partial products have been retained, we generate the sum.  In this example in step \textcircled{1} $T = A + A<<2 + A<<3 + A<<4 + A<<6$, and in step \textcircled{2} product $P=T+A<<9 + A<<10+A<<11+A<<14$.  Again, this takes only two addition steps.  In the worst case, this takes $\frac{n}{\text{TRD}-2}n$ steps or $O(n^2)$ complexity where $n$ is the bit-width of the operands.

\begin{algorithm}[t]
\DontPrintSemicolon
\footnotesize
\KwData{Two rows A and B of size $n$ with word of size $w$}
\KwResult{S=A*B}
\Begin{
Inter-bank-copy(A) $\rightarrow$ 0th position of target DBC\;
\For{$i \in 0..w-1$}{
Read-shifted\;
Write\;
DW Shift right\;
}\;
Precharge write driver to ``0..0''\;
Inter-bank-copy(B) $\rightarrow$ local-row-buffer\;
\For{$i \in w-1..0$}{
  \For{$j \in 0..n/2w-1$} {
    \If{$b_{j*2w+i} = 0$}{write $b_{j*2w}..b_{(j+1)2w-1}$}\; 
  }
DW Shift left\;

}\;

\For{$i \in w-1..0$, $i+=$TRD} {
Compute S, C and C'\;
Shift DBC + TRD\;
Write S, C, C' in DBC'\;
Shift DBC' + 3\;
}

Compute S, C, and C' from DBC' $\rightarrow$ DBC\;
Continue until $\leq TRD-2$ elements\;
Add to compute final result\;
}
\caption{Parallel Multiplication}
\label{Algo}
\end{algorithm}

\subsubsection{Optimized Multiplication}
\label{sub_sec:ExampleOfOptimized Multiplication}
\label{sec:optimization}

In the prior methods, we required several addition steps to generate the final result and each addition is still performed sequentially.  Using the partial products methodology the complexity remains $O(n^2)$ in the bit-width.  To improve on this we can borrow from Carry Save Adders (CSAs).  A CSA leverages the three inputs of a full adder $A,B,C_{IN}$ to be used for three operands $X,Y,Z$ instead of two.  This creates an entirely parallel process to reduce three operands to two in the form of $S^\dagger,C^\dagger$.  Then a traditional addition using a ripple carry adder can add $S^\dagger+C^\dagger \rightarrow S$.  We can leverage our polymorphic gates in the same way but with more input operands where seven partial products can be reduced to a $S,C,C'$ in parallel yielding a $7\rightarrow3$ operand reduction function.  

This method accomplishes two things.  The need for the sequential carry logic of addition is not required for the reduction step and the technique can directly be performed on TRD instead of $\text{TRD}-2$ operands. 
Furthermore, the $7\rightarrow3$ reduction operation can be repeated on data, including prior output of this reduction function in a previous step.  These reductions are continued until there are $\leq (\text{TRD}-2)$ operands remaining.  The final result can then be computed with a single addition operation, where one output is generated by TRD-2 inputs.  For 32-bit multiplication in the worst case this would require $n-1$ logical shifts, seven $7\rightarrow3$ transformations, and one add that requires $n$ steps.  This makes multiply an $O(n)$ operation.

The power of this approach is that the number and latency of $7 \rightarrow 3$ transformations is much less than the latency of the add operation.  Thus, given the width of the data stored in the DBC as 512, we can pack 32-bit multiply operations into the row width and gain parallelism while retaining simple control that allows the multiply to execute while only blocking a single subarray, allowing other subarrays to proceed in parallel.  This general multiplication leveraging CSA inspired $7 \rightarrow 3$ reduction of multiple packed $w$-bit words per $n$-bit row is specified in Algorithm~\ref{Algo}.
Note, as multiplication increases a $2w$-bit wide product, we pack $w$-bit wide words with $w$-bit wide gaps to store for the full product.

First, one operand is copied across banks to a processing tile as described in prior work~\cite{seshadri2013rowclone}.
Once A is in the 0th position of a DBC, the first loop shifts and writes the row $w$ times in adjacent domains.  Then we precharge the write driver to ``0..0'' and bring the B row into the rowbuffer.  Next, we shift back along the DBC and zero out the $ith$ shifted version of A if that bit of B is `0'.  We do this for each of the packed words in the row independently.  This is the only portion of the control that differs based on the different words and it functions like predicated execution.  We have now returned to the origin and start to use PIM to compute $S,C,C'$ over the TRD and write this into another DBC or DBC$'$.  The remaining partial products from DBC are converted into $S,C,C'$ in DBC$'$.  Then DBC$'$ is traversed writing the newly computed $S,C,C'$ back to the original DBC.  This continues until there are $\leq\text{TRD}-2$ elements, which triggers an add to compute the final result.  
There are a few minor optimizations to this flow that can minimize cycles and steps, such as interleaving shifted versions of A across two DBCs in groups of size TRD which can overlap shifting and computing.  Also for fixed size $w,n,$TRD, the exact steps can be optimized.  

Another possible optimization for large addition is to increase TRD, consequently the number of operands per operation. For 8$\leq$TRD$\leq$15, in addition to $S,C,C'$, PIRM could generate $C''$ that would impact the bitline+3. 

\section{Case Study: Implementing a CNN}
\label{subSec:CNNUsingPIRM}

To demonstrate the potential of PIRM we explore the process of computing a convolutional neural network (CNN) using PIM.  The CNN approach is composed of 3 main types of layers: convolution layers, pooling layers and fully connected layers, each of which can be completed in PIRM.
\subsection{Convolution}
\newcommand{\myMatrix}[1]{\mathbf{{#1}}}

The convolution layer is the process of taking a small kernel (or weight) matrix $\myMatrix{K}$, and combining it in ``windows'' with a larger matrix $\myMatrix{I}$ representing input features, at each step multiplying the overlapping positions and accumulating the results. As an example, for $\myMatrix{I}$ and $\myMatrix{K}$ of size $NxN$ and $3x3$, respectively, the convolution operation is:

\begin{equation}
    Conv(\myMatrix{I},\myMatrix{K})(m,p)= \sum_{j=0}^{2} \sum_{t=0}^{2}  \myMatrix{K}_{j,t}*\myMatrix{I}_{m+j-1,p+t-1}
\end{equation}

\begin{figure}[btp]
    \centering
    \includegraphics[width=\columnwidth]{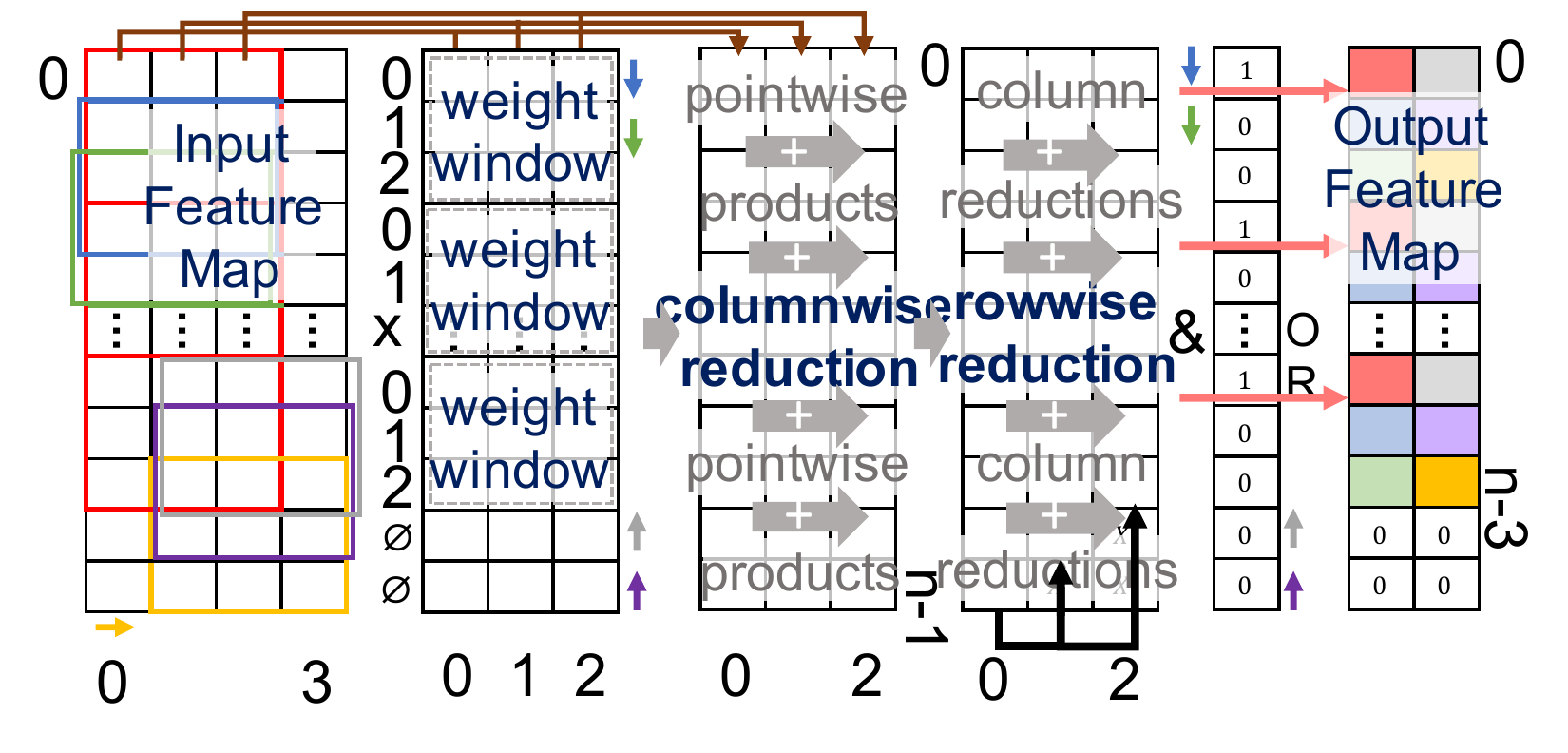}
    \caption{Example convolution flow using PIRM}
    \label{fig:convolution}
\end{figure}

Using PIRM, we show the convolution flow in Fig.~\ref{fig:convolution}.  First the multiplications are completed by operating in parallel with multiple copies of the weight window and executed on multiple items packed into rows in parallel.  The multiplication  approach in Section~\ref{sub_sec:ExampleOfOptimized Multiplication} generates $S,C,C'$ in parallel and summed with the 5-operand add  
as described in Section~\ref{sub_sec:addition}. This generates the pointwise multiplication results.  These results are then reduced over summation first across the columns, then those results are summed across the rows by shifting them into alignment to generate the output features.  For larger kernels, PIRM performs  $7\rightarrow3$ reductions as needed.

\subsection{Pooling}

During pooling, the dimensionality of an input matrix is reduced by taking the average or maximum of all values in submatrices of a predefined size to generate the output matrix values. 
Using PIRM, the max function can be realized via TR across all the candidates evaluating MSB to LSB sequentially.  Each step compares the binary weight of the same bit position in each word, and the TR result determines the subsequent action via predicated execution with local information.  This allows PIM instructions issued by the memory controller to work in parallel across many subarrays in parallel.  
First, the values upon which to compute the max are stored in adjacent positions between the access points. Then 
a TR is performed across the MSBs; If TR$>$0 the value under the right head is read and stored in the rowbuffer. If the MSB is `0' the rowbuffer is reset.  This eliminates a value that is lower than the other values.  Then the DBC is shifted right and the value of the rowbuffer is written to the left head.  

All five words are processed in this manner.  If TR $=0$, then each word is read from the right and re-written to the left access point, while shifting in between.  Essentially, the data remains unchanged.  This is necessary because if all values are `0' in this position, it does not eliminate any values from being the maximum.  From a PIM instruction execution perspective, the memory controller issued instructions are identical for all participating subarrays by making the rowbuffer reset command predicated on the TR and tested bit.  We can use a DWM $\texttt{AND}$ function in another DBC of the same tile/subarray to store and compute the logic value governing the predicated execution of the row-buffer reset.  

\begin{figure}[t]
 \centering
\includegraphics[width=0.8\columnwidth]{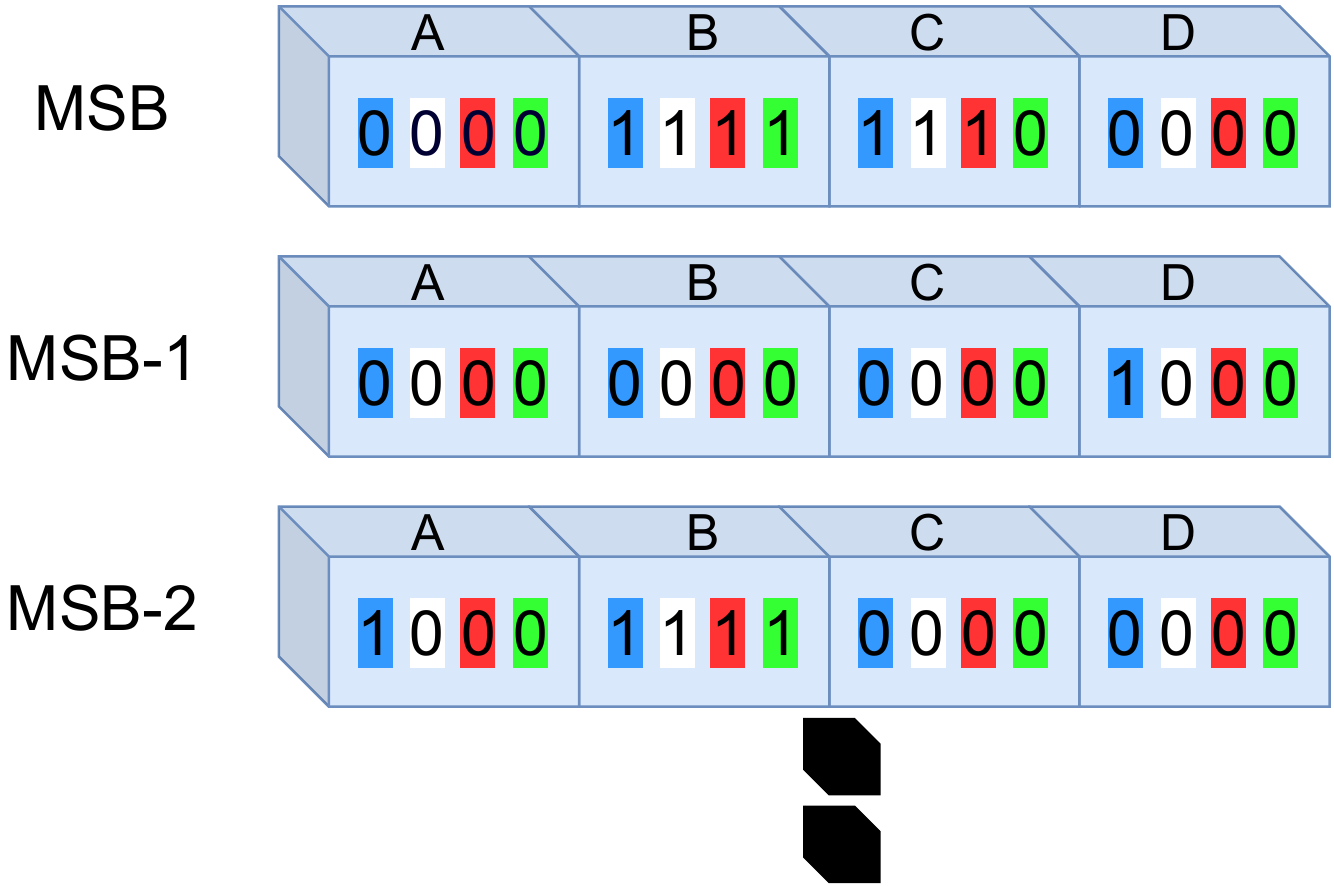}
\caption{Maximum Function Example}
\label{fig:Maximum}
\end{figure}

\begin{figure}[b]
 \centering
\includegraphics[width=\columnwidth]{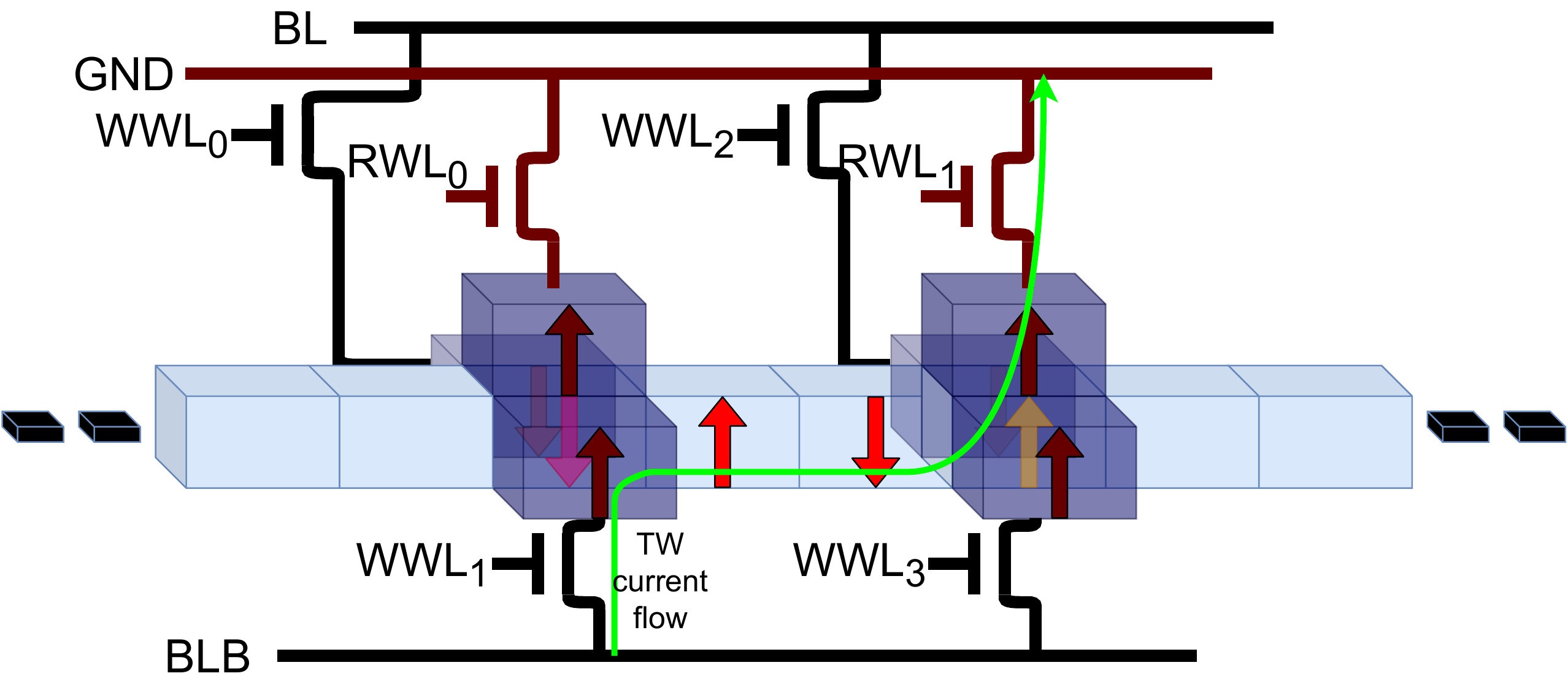}
\caption{Transverse write and segmented shift.}
\label{fig:TW}
\end{figure}

The process is repeated for each bit position and the value is read using TR$>0$, so the max vector is read, regardless of its location between the heads and if $>1$ vectors equal the max value the TR value it is still accessed correctly.  
Fig.~\ref{fig:Maximum} depicts a concrete example of the state of four words A, B, C, and D as they are processed by the maximum subroutine, in chronological order from left to right with a different color representing the values after a bit is processed.  At the MSB pass starting with blue, TR$>0$.  Words A and D have `0' in their MSB, so they are overwritten by the zero vector and B and C are written back unchanged.  The result after the first step is shown in white.  
For MSB-1, TR $=0$.  Thus all words are read and written back unchanged as shown in red. For MSB-2, TR $>0$.  Words A, C, and D all have `0' in that bit position, so they are overwritten by the zero vector and only B is written back unchanged as shown in green.  Now the maximum value has been determined, and will be maintained as all the bits are traversed through the LSB. 
The maximum function requires cycling through vectors many times which makes shifting the entire nanowire impractical.  To address this concern, and to reduce delay, we propose novel technique Transverse Write (TW) technique with \textit{segmented shifting}.  This is inspired by the shift-based writing~\cite{DWM_Tapestri} approach and transverse access techniques~\cite{roxy2020novel} previously proposed.  We illustrate this new concept in Fig.~\ref{fig:TW}.  Two write/read heads are represented in dark blue separated by data domains in light blue (four domains are shown instead of seven to simplify the explanation). 
To perform a classic shift-based write under the left head, \texttt{WWL}$_1$ and \texttt{RWL}$_0$ are closed, thus the current flows from \texttt{BLB} to a fixed layer, to the domain, shifting the dark red upward orientation of the fixed layer to replace the pink downward orientation in the nanowire. 
This operation can be modified to shift the pink orientation along the nanowire rather than to ground.
This TW operation closes 
\texttt{WWL}$_1$ and \texttt{RWL}$_1$. Thus, the current flows from \texttt{BLB} through the fixed layer and the four domains before exiting through the right head as indicated by the green arrow. By doing so, the fixed layer orientation at \texttt{WWL}$_1$ is written under the left head, and the pink orientation and those which follow it advance along the nanowire, forcing the yellow arrow to \texttt{GND}.  

Applying this to the majority function, in the context of Fig.~\ref{fig:TW}, if TR $>0$, PIRM reads from the right head (yellow arrow).
The predicated rowbuffer reset command is executed.  Then the value of the word is written via TW from the left head to the right head.  Thus, by reading from the right and conducting TW from the left the segmented shift ensures each updated operand is returned to its original position along the nanowires and the remaining locations are not disturbed.

\subsection{Fully Connected}
The fully connected layer executes the following function:

\begin{equation}
    ReLU(\myMatrix{W}\myMatrix{x}+\myMatrix{b})
    \label{eq:fully-connected}
\end{equation}

\noindent where $\myMatrix{W}$ is the weight matrix, $\myMatrix{x}$ is the input vector and $\myMatrix{b}$ is the bias vector. The $ReLU$ function returns zero if $\sum_{i=0}^{i_0} \myMatrix{W}_{ij}\myMatrix{x}_i+\myMatrix{b}_j\leq0$ and $\sum_{i=0}^{i_0} \myMatrix{W}_{ij}\myMatrix{x}_i+\myMatrix{b}_j$ otherwise. This function is implemented by computing $\sum_{i=0}^{i_0} \myMatrix{W}_{ij}\myMatrix{x}_i+\myMatrix{b}_j$ using PIRM addition and multiplication operations.  Using a predicated row refresh based on the MSB, which is `1' for values $<0$ and writing the value back the resulting value from the $ReLU$ function to the array, repeating $\forall j$.

\section{Experimental Results}
\label{sec:experimentalResult}

\begin{table*}[t]
\centering
\caption{Operation Comparison }
\label{tab:Performances}
\resizebox{\linewidth}{!}{
\begin{tabular}{l|ccc|cccc|cccc}
\hline\hline
\textbf{Scheme} & \multicolumn{3}{|c|}{\textbf{PIRM}} & \multicolumn{4}{|c|}{\textbf{DW-NN}} & \multicolumn{4}{|c}{\textbf{SPIM}} \\
\textbf{Unit} & \textbf{2op Add} & \textbf{5op Add} & \textbf{Mult} & \textbf{2op Add} & \textbf{5op Add} & \textbf{5op Add} & \textbf{2op Mult.} & \textbf{2op Add} & \textbf{5op Add} & \textbf{5op Add} & \textbf{2op Mult.} \\
 & (TR = 4) & (TR = 7) & (TR = 7) &  & Area Opt. & Lat. Opt. &  &  & Area Opt. & Lat. Opt. &  \\\hline
\textbf{Speed} (cycles) & 26 & 26 & 64 & 54 & 264 & 194 & 163 & 49 & 244 & 179 & 149 \\
\textbf{Energy} (pJ) & 12.54 & 22.14 & 57.39 & 40 & 169.6 & 169.6 & 308 & 28 & 121.6 & 121.6 & 196 \\
\textbf{Area} ($\mu m^{2}$) & 2.16 & 4.94 & 5.07 & 2.6 & 2.6 & 5.2 & 18.9 & 2 & 2 & 4 & 16.8\\\hline\hline
\end{tabular}
}
\end{table*}

We present four experiments to demonstrate the effectiveness of PIRM. First, we compare  
PIRM addition and multiplication characteristics against other computing units based on DWM. Second, we demonstrate the benefit of PIRM PIM on addition/multiplication oriented benchmarks from polybench~\cite{pouchet2012polybench}.
Third, we compare bitmap indices~\cite{chan1998bitmap}, a common component of database queries, against Ambit and ELP$^2$IM to show the benefit over state-of-the-art DRAM PIM.  Finally, we implement two CNN applications, Lenet5 and Alexnet, using the method in Section~\ref{subSec:CNNUsingPIRM}, and compare PIRM to SPIM, Ambit, and ELP$^2$IM.

\subsection{Comparison with DWM PIM technique}
\label{sec:compDWMPIMTechnique}

Based on the device level information provided in~\cite{roxy2020novel, nvsim,nvsimRT, yu2014energy}, we calculated the timing and energy of read, write and TR operations for DWM. We have designed PIRM's sense circuits for TR and synthesized the PIM logic from Fig.~\ref{fig:PIMgates} in 45nm technology using FreePDK45~\cite{FreePDK45} and the Cadence Encounter flow.  We then scaled the design by normalizing $F^2$ to 32nm as described in prior work~\cite{nvsim, chen2010advances} to compare with the 32nm results reported in DW-NN~\cite{yu2014energy}.
To calculate the energy we used a modified version of NVSIM to report the DWM energy at 32nm and modified the SA energy using our custom sensing circuit designed in LTSPICE and scaled energy reported from ASIC synthesis for the PIM logic gates.
Latency is obtained by calculating the number of operations needed to perform an 8-bit add or multiply operation presuming a 1ns cycle speed, consistent with values reported by NVSIM.  PIRM 8-bit addition shifts and writes the words between the two heads (10 cycles) and then writes after each TR (16 cycles) which yields to a total of 26 cycles.  Table~\ref{tab:Performances} reports the speed, energy, and area of PIRM, DW-NN and SPIM for two operand addition (2op add), five operand addition optimized for area by conducting multiple additions in series (5op add area), five operand addition optimized for latency by replicating addition units (5op add latency), and two operand multiplication (2op mult).  PIRM is 1.9$\times$, 9.4$\times$, 6.9$\times$ and 2.3$\times$ faster and 2.2$\times$, 5.5$\times$, 5.5$\times$ and 3.4$\times$ less energy than SPIM, the state-of-the-art technique, for 2op add, 5op add area optimized, 5op add latency optimized, and 2op multiplication, respectively.  PIRM is comparable to DW-NN and requires some area increase over SPIM for addition, but reduces the multiplication area by 3.7$\times$ and 3.3$\times$ compared to DW-NN and SPIM, respectively, while also providing additional processing capabilities such as bulk-bitwise operations. 

\subsection{Improvement to Memory Wall Versus Non-PIM DWM}
\label{sub_sec:ImprovementOverDWM}

We tested our scheme on the standard polyhedral polybench using a modified version of RTSIM~\cite{khan2019rtsim}.
The Polybench consists of 29 applications from different domains including linear algebra, data mining, and stencil kernels. From these 29 applications we selected the benchmarks most heavily focused on matrix addition and multiplication, from \texttt{2mm}, which is two matrix multiplication, to \texttt{gemm} which is governed by $\myMatrix{C}=\alpha\myMatrix{A}\myMatrix{B}+\beta\myMatrix{C}$. 
We provide energy and queue latency improvement and the area overhead of PIRM at the main memory granularity. For this simulation we used the parameters of Table~\ref{tab:Parameters}~\cite{ molka2010characterizing}.

\subsubsection{Queue Latency}
\label{sub_sec:QueueLatency}

\begin{figure}[b]
 \centering
\includegraphics[width =\columnwidth]{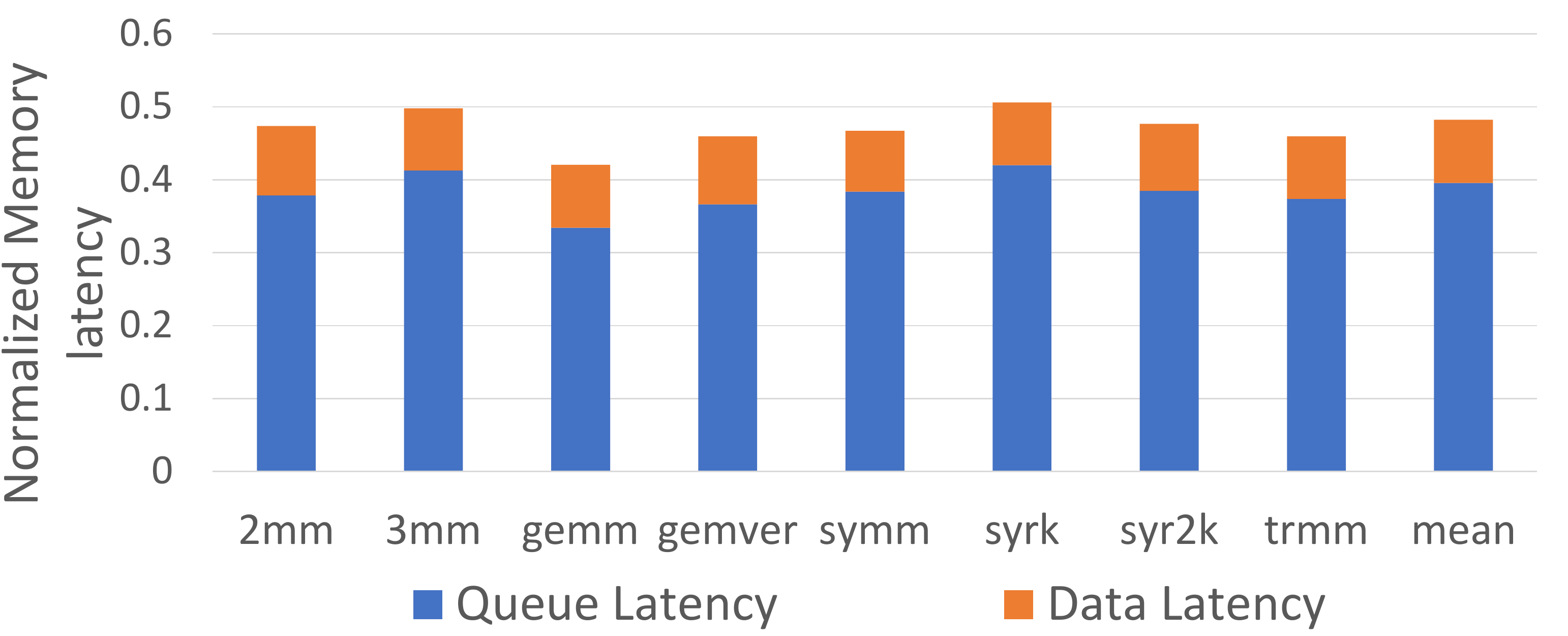}
\caption{Normalized DWM Latency}
\label{fig:latency2}
\end{figure}

In these experiments, we extract the traces using a pintool, then we examine the accesses and determine which accesses correspond to additions and multiplications and we determine if there is an available PIRM PIM-enabled tile to perform the operations in memory instead of sending it to the processing unit. In order to efficiently use our PIM, we issue multiple parallel arithmetic instructions from the memory controller to different PIM tiles, which allows single instruction multiple data (SIMD) operation.
Over the benchmarks shown in Fig.~\ref{fig:latency2} we demonstrate an average improvement over the memory latency of 2.1$\times$.

\begin{table}[t]
\centering
\caption{DWM parameters}
\label{tab:Parameters}
\begin{tabular}{|l|l|l|l|}
\hline
Memory size &  8GB & Processor & Intel Xeon X5670 \\
Number of Bank & 32 & $E_{trans}$ & 1250 (pJ/Byte) \\ 
Subarrays per Bank & 64 & addition (32 bits) & 111 (pJ/op)\\ 
Tile per Subarray & 16 & multiplication & 164 (pJ/op)\\ 
DBC per Tile & 16 & &  \\\hline

\end{tabular}
\end{table}

\subsubsection{Area overhead}
\label{sub_sec:AreaOverhead}

\begin{figure}[t]
 \centering
\includegraphics[width =\columnwidth]{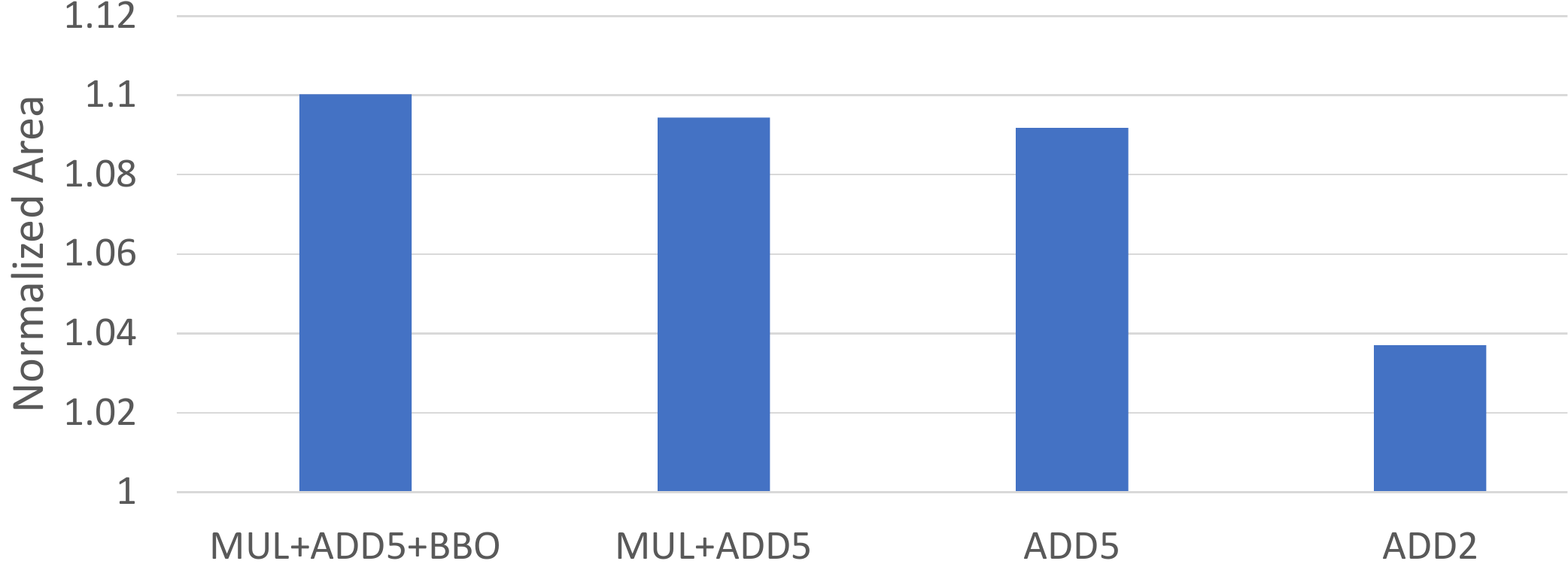}
\caption{Area overhead over DWM memory for four PIM designs, multiplication plus addition plus bulk bitwise operations, multiplication and addition of five words, addition of 5 words only and finally addition of 2 words}
\label{fig:area2}
\vspace{-.1in}
\end{figure}

Fig.~\ref{fig:area2} shows the area overhead for adding some PIM capability to one tile in each subarray of the memory.  There is a 10\% overhead for including our full PIM ISA including multiplication, five operand addition, and seven operand bulk-bitwise operations.  By stripping the bulk-bitwise operations, this overhead reduces to about 9\%, removing the multiplication also keeps us around 9\%.  Dropping from a five to two operand adder reduces the overhead to $<4\%$.  Another way to reduce overhead is to interleave PIM units across only half the subarrays to get an overhead of around 5\%, or every fourth subarray to drop it to 2.5\%.  Even at this 2.5\% overhead, an 8GB PIRM memory would still have 8192 PIM enabled DBCs and can still operate as a highly parallel SIMD PIM engine.

\begin{figure}[t]
 \centering
\includegraphics[width=\columnwidth]{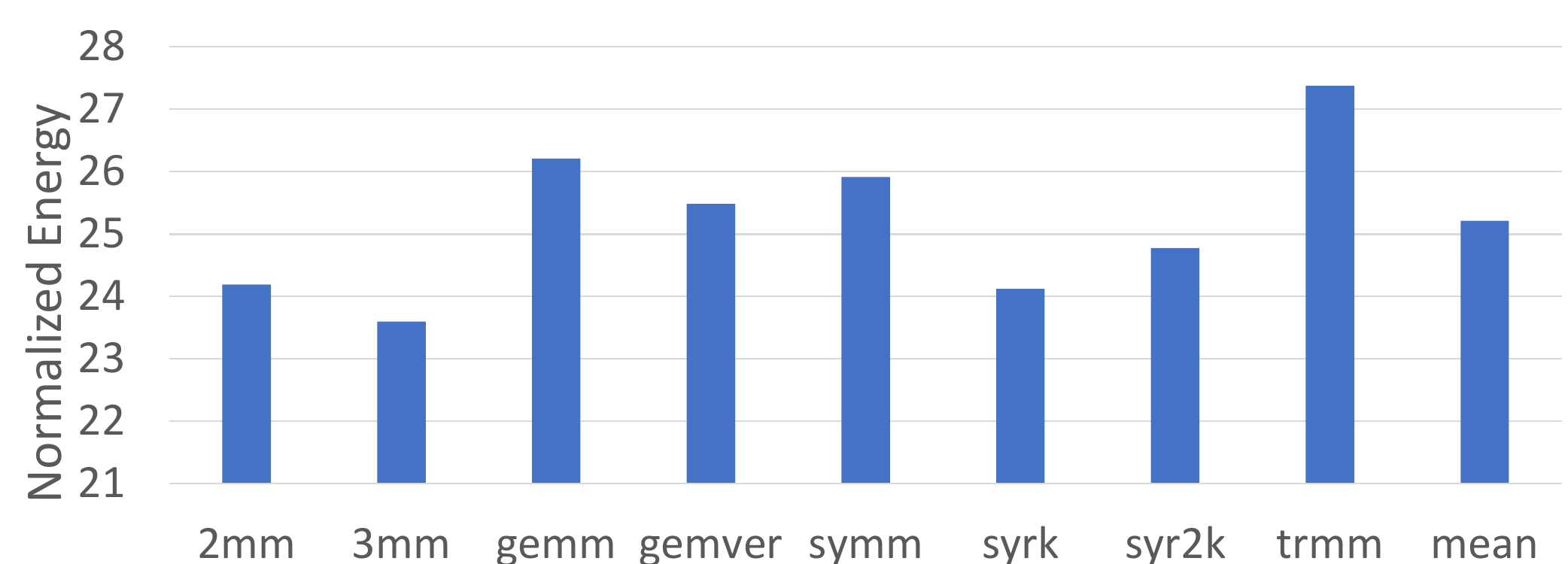}
\caption{Energy Improvement over Polybench benchmark}
\label{fig:energy}
\end{figure}

\subsubsection{Energy consumption}
\label{sub_sec:energyConsumption}

To perform an energy analysis we compare the energy to send the data from DWM to the CPU and back plus the energy required to perform a CPU operation to the energy that PIRM needs to perform the same operation. The data transfer and operation costs are listed in Table~\ref{tab:Parameters}.  While the energy to perform the operation is the same order of magnitude in the CPU and in PIRM, the price to send the data is extremely high, more than 30$\times$ the operation cost. Leveraging these savings, Fig.~\ref{fig:energy} reports the energy improvement over the polybench benchmarks when leveraging PIM as much as possible, which decreases the energy consumption by more than 25.2$\times$, on average.

\subsection{Bitmap Indices}
\label{sub_sec:bitmap}

While PIRM has significant benefits over SPIM and DW-NN, neither of these schemes can perform bulk-bitwise logic.  Thus, we compare the bulk-bitwise capabilities of PIRM to the state-of-the-art technique for bulk-bitwise operation in DRAM, ELP$^2$IM and Ambit.  We make this comparison on a classic PIM application as an indication of what PIRM could achieve against a scheme similar to a currently deployed memory technology.

\begin{figure}[b]
 \centering
\includegraphics[width =.8\columnwidth]{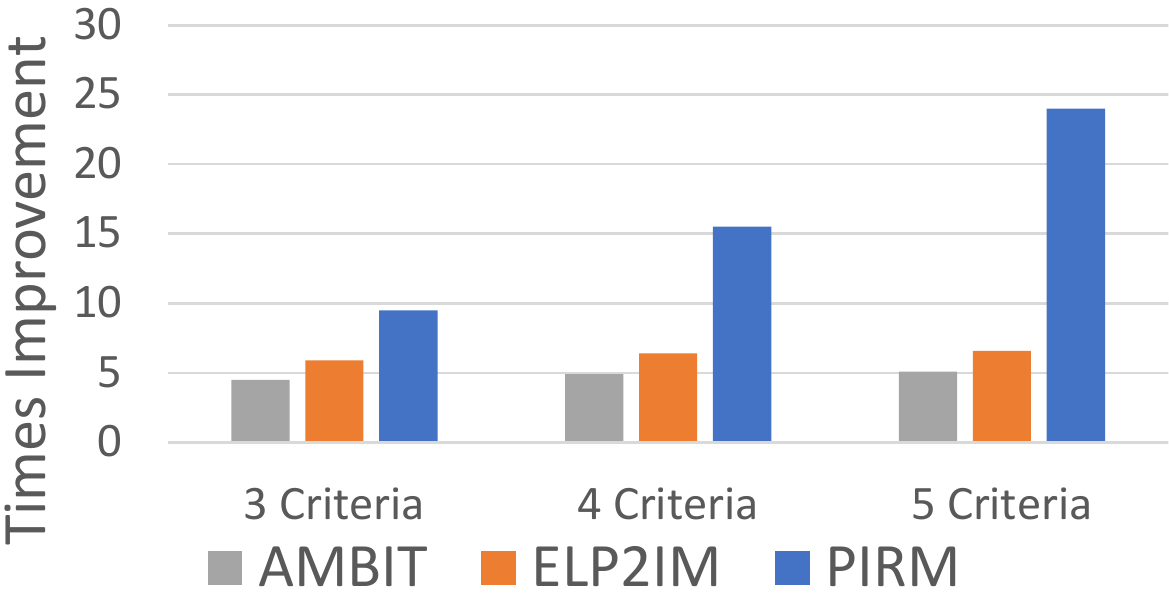}
\caption{Normalized performance}
\label{bitwiseperf}
\label{fig:bitwiseperf}
\end{figure}

We repeated the bitmap indices database query~\cite{chan1998bitmap} experiment from prior DRAM PIM work~\cite{elp2im}.  A query from 16 million users' data requested how many male users were active in the past $w$ weeks, where $w \in \{2..4\}$.  Fig.~\ref{fig:bitwiseperf} shows the latency improvement of Ambit, ELP2IM and PIRM normalized to the latency of a standard DRAM CPU system.  For three, four, and five search criteria, \textit{i.e.,} male users for last two, three, and four weeks, PIRM provides a 1.6$\times$, 2.2$\times$, and $3.4\times$ query speedup, respectively over the state-of-the-art ELP$^2$IM.  The speedup achieved when including more search criteria demonstrates the benefit of PIRM given its multi-operand bulk bitwise operations compared to two operand limits of previous work.

\subsection{CNN}
\label{sec:convolution}

\begin{table}[t]
\centering
\caption{CNN application comparison}
\label{tab:CNNComp}
\resizebox{0.7\columnwidth}{!}{
\begin{tabular}{l|c|c}
\hline\hline
\textbf{Scheme} & {\textbf{Alexnet}} & {\textbf{Lenet5}}
\\\hline
\textbf{SPIM} (FPS) & 22.5 & 53.2\\
\textbf{Ambit-NID} (FPS) & 84.8 & 7697.4\\
\textbf{ELP2IM-NID} (FPS) & 96.4 & 8329.5 \\
\textbf{PIRM} (FPS) & 5217.9 & 130393.4 \\\hline
\textbf{Improvement} & 54$\times$ & 15$\times$ \\\hline\hline
\end{tabular}
}

\end{table}

We implemented two CNN benchmarks: CNN Lenet-5~\cite{lenet} and Alexnet~\cite{alexnet}, commonly used for image processing, machine learning training on handwritten digits and on RGB images.  We used the parameters of Table~\ref{tab:Parameters} with a clock frequency of 1GHz.  First, we demonstrate the benefit of PIRM over DW-NN and SPIM for convolution (Fig.~\ref{fig:convolution}).  Input features and weights can be packed into memory rows and combined with point-wise multiplication.  We assume similar packing is possible in SPIM and DW-NN.  Where PIRM really shines is in the addition reduction operations.  By leveraging multi-operand addition, logical shifting, and a second multi-operand reduction, PIRM can reduce up to a 5x5 convolution window, common in both algorithms, in two addition steps.  Presuming the same memory configuration from Table~\ref{tab:Parameters} for all schemes, PIRM is capable of executing convolution at 26 Tera Ops Per Second (TOPS), while DW-NN and SPIM achieve 2.88 and 2.92 TOPS, respectively, for the same number of functional units.  PIRM is also capable of 108 Giga Ops Per Joule (GOPJ) while DW-NN and SPIM achieve 17 and 25 GOPJ, respectively.  For context, a dedicated FPGA CNN accelerator was able to achieve 0.34 TOPS with 12.5 GOPJ~\cite{jiang2019achieving}.   

We implemented the full CNNs, including the pooling and fully connected layers, as described in Section~\ref{sec:convolution}. Table~\ref{tab:CNNComp} shows PIRM improvement over the state-of-the-art PIM techniques in DWM, SPIM, and in DRAM, Ambit and ELP2IM applied to NID~\cite{sim2018nid}. 
PIRM executes 54$\times$ and 15$\times$ more Frames Per Second (FPS) compared to ELP$^2$IM which was already 1.22$\times$ faster than Ambit.

\section{Conclusion}
\label{sec:conclusion}

PIRM is a significant step forward for PIM in the promising DWM technology. Our technique takes advantage of the intrinsic proximity of bits in DWM nanowires and the advantages of TR to build a polymorphic gate that can support myriad PIM operations.  PIRM can perform bulk-bitwise or addition on multiple operands simultaneously, limited only by the TRD between access ports on DWM.  Using carry-select adder inspired techniques these multi-operand operations can be used to efficiently implement multiplication with minimal additional logic.  Our results show that PIRM improves over the state-of-the-art DWM-based PIM by 6.9$\times$ and 2.3$\times$ in term of speed and 5.5$\times$ and 3.4$\times$ in terms of energy for five operand addition (optimized for latency) and multiplication, respectively. Thus, in terms of a convolution application, PIRM can perform 8.9$\times$ and 4.3$\times$ more GOPS and GOPJ, respectively.
Compared to a standard DWM memory without PIM, PIRM improves memory latency by 2.1$\times$, decreases  energy by 25.2$\times$ versus sending the data to the CPU.  PIRM incurs an area overhead of 10\% when PIM enabling one tile per subarray.  PIRM bulk-bitwise capabilities are $\geq1.6\times$ faster than the state-of-the-art DRAM approach, ELP$^2$IM.  Moreover, PIRM provides a superset of the existing PIM techniques in a single memory technology.

\footnotesize
\bibliography{main}

\begin{thebibliography}{10}
\providecommand{\url}[1]{#1}
\csname url@samestyle\endcsname
\providecommand{\newblock}{\relax}
\providecommand{\bibinfo}[2]{#2}
\providecommand{\BIBentrySTDinterwordspacing}{\spaceskip=0pt\relax}
\providecommand{\BIBentryALTinterwordstretchfactor}{4}
\providecommand{\BIBentryALTinterwordspacing}{\spaceskip=\fontdimen2\font plus
\BIBentryALTinterwordstretchfactor\fontdimen3\font minus
  \fontdimen4\font\relax}
\providecommand{\BIBforeignlanguage}[2]{{%
\expandafter\ifx\csname l@#1\endcsname\relax
\typeout{** WARNING: IEEEtran.bst: No hyphenation pattern has been}%
\typeout{** loaded for the language `#1'. Using the pattern for}%
\typeout{** the default language instead.}%
\else
\language=\csname l@#1\endcsname
\fi
#2}}
\providecommand{\BIBdecl}{\relax}
\BIBdecl

\bibitem{villa2014scaling}
O.~Villa, D.~R. Johnson, M.~Oconnor, E.~Bolotin, D.~Nellans, J.~Luitjens,
  N.~Sakharnykh, P.~Wang, P.~Micikevicius, A.~Scudiero \emph{et~al.}, ``Scaling
  the power wall: a path to exascale,'' in \emph{SC'14: Proceedings of the
  International Conference for High Performance Computing, Networking, Storage
  and Analysis}.\hskip 1em plus 0.5em minus 0.4em\relax IEEE, 2014, pp.
  830--841.

\bibitem{mckee2004reflections}
S.~A. McKee, ``Reflections on the memory wall,'' in \emph{Proceedings of the
  1st conference on Computing frontiers}, 2004, p. 162.

\bibitem{molka2010characterizing}
D.~Molka, D.~Hackenberg, R.~Sch{\"o}ne, and M.~S. M{\"u}ller, ``Characterizing
  the energy consumption of data transfers and arithmetic operations on x86- 64
  processors,'' in \emph{International conference on green computing}.\hskip
  1em plus 0.5em minus 0.4em\relax IEEE, 2010, pp. 123--133.

\bibitem{elp2im}
X.~Xin, Y.~Zhang, and J.~Yang, ``Elp2im: Efficient and low power bitwise
  operation processing in dram,'' in \emph{2020 IEEE International Symposium on
  High Performance Computer Architecture (HPCA)}.\hskip 1em plus 0.5em minus
  0.4em\relax IEEE, 2020, pp. 303--314.

\bibitem{seshadri2017ambit}
V.~Seshadri, D.~Lee, T.~Mullins, H.~Hassan, A.~Boroumand, J.~Kim, M.~A. Kozuch,
  O.~Mutlu, P.~B. Gibbons, and T.~C. Mowry, ``Ambit: In-memory accelerator for
  bulk bitwise operations using commodity dram technology,'' in \emph{2017 50th
  Annual IEEE/ACM International Symposium on Microarchitecture (MICRO)}.\hskip
  1em plus 0.5em minus 0.4em\relax IEEE, 2017, pp. 273--287.

\bibitem{li2016pinatubo}
S.~Li, C.~Xu, Q.~Zou, J.~Zhao, Y.~Lu, and Y.~Xie, ``Pinatubo: A
  processing-in-memory architecture for bulk bitwise operations in emerging
  non-volatile memories,'' in \emph{Proceedings of the 53rd Annual Design
  Automation Conference}, 2016, pp. 1--6.

\bibitem{yu2014energy}
H.~Yu, Y.~Wang, S.~Chen, W.~Fei, C.~Weng, J.~Zhao, and Z.~Wei, ``Energy
  efficient in-memory machine learning for data intensive image-processing by
  non-volatile domain-wall memory,'' in \emph{2014 19th Asia and South Pacific
  Design Automation Conference (ASP-DAC)}.\hskip 1em plus 0.5em minus
  0.4em\relax IEEE, 2014, pp. 191--196.

\bibitem{CNN_DWM}
B.~Liu, S.~Gu, M.~Chen, W.~Kang, J.~Hu, Q.~Zhuge, and E.~H.-M. Sha, ``An
  efficient racetrack memory-based processing-in-memory architecture for
  convolutional neural networks,'' in \emph{2017 IEEE International Symposium
  on Parallel and Distributed Processing with Applications and 2017 IEEE
  International Conference on Ubiquitous Computing and Communications
  (ISPA/IUCC)}.\hskip 1em plus 0.5em minus 0.4em\relax IEEE, 2017, pp.
  383--390.

\bibitem{li2017drisa}
S.~Li, D.~Niu, K.~T. Malladi, H.~Zheng, B.~Brennan, and Y.~Xie, ``Drisa: A
  dram-based reconfigurable in-situ accelerator,'' in \emph{2017 50th Annual
  IEEE/ACM International Symposium on Microarchitecture (MICRO)}.\hskip 1em
  plus 0.5em minus 0.4em\relax IEEE, 2017, pp. 288--301.

\bibitem{subramaniyan2017parallel}
A.~Subramaniyan and R.~Das, ``Parallel automata processor,'' in \emph{2017
  ACM/IEEE 44th Annual International Symposium on Computer Architecture
  (ISCA)}.\hskip 1em plus 0.5em minus 0.4em\relax IEEE, 2017, pp. 600--612.

\bibitem{wong2010phase}
H.-S.~P. Wong, S.~Raoux, S.~Kim, J.~Liang, J.~P. Reifenberg, B.~Rajendran,
  M.~Asheghi, and K.~E. Goodson, ``Phase change memory,'' \emph{Proceedings of
  the IEEE}, vol.~98, no.~12, pp. 2201--2227, 2010.

\bibitem{luo2018write}
H.~Luo, Q.~Liu, J.~Hu, Q.~Li, L.~Shi, Q.~Zhuge, and E.~H.-M. Sha, ``Write
  energy reduction for pcm via pumping efficiency improvement,'' \emph{ACM
  Transactions on Storage (TOS)}, vol.~14, no.~3, pp. 1--21, 2018.

\bibitem{song2018graphr}
L.~Song, Y.~Zhuo, X.~Qian, H.~Li, and Y.~Chen, ``Graphr: Accelerating graph
  processing using reram,'' in \emph{2018 IEEE International Symposium on High
  Performance Computer Architecture (HPCA)}.\hskip 1em plus 0.5em minus
  0.4em\relax IEEE, 2018, pp. 531--543.

\bibitem{yao2020fully}
P.~Yao, H.~Wu, B.~Gao, J.~Tang, Q.~Zhang, W.~Zhang, J.~J. Yang, and H.~Qian,
  ``Fully hardware-implemented memristor convolutional neural network,''
  \emph{Nature}, vol. 577, no. 7792, pp. 641--646, 2020.

\bibitem{kang2017memory}
W.~Kang, H.~Wang, Z.~Wang, Y.~Zhang, and W.~Zhao, ``In-memory processing
  paradigm for bitwise logic operations in stt--mram,'' \emph{IEEE Transactions
  on Magnetics}, vol.~53, no.~11, pp. 1--4, 2017.

\bibitem{Parkin-08-Science}
S.~S.~P. Parkin, M.~Hayashi, and L.~Thomas, ``Magnetic domain-wall racetrack
  memory,'' \emph{Science}, vol. 320, no. 5874, pp. 190--194, Apr. 2008.

\bibitem{roxy2020novel}
K.~Roxy, S.~Ollivier, A.~Hoque, S.~Longofono, A.~K. Jones, and S.~Bhanja, ``A
  novel transverse read technique for domain-wall “racetrack” memories,''
  \emph{IEEE Transactions on Nanotechnology}, vol.~19, pp. 648--652, 2020.

\bibitem{zhang2012perpendicular}
Y.~Zhang, W.~Zhao, D.~Ravelosona, J.-O. Klein, J.-V. Kim, and C.~Chappert,
  ``Perpendicular-magnetic-anisotropy cofeb racetrack memory,'' \emph{Journal
  of Applied Physics}, vol. 111, no.~9, p. 093925, 2012.

\bibitem{DWM_Tapestri}
R.~Venkatesan, M.~Sharad, K.~Roy, and A.~Raghunathan, ``Dwm-tapestri-an energy
  efficient all-spin cache using domain wall shift based writes,'' in
  \emph{Proc. of DATE}, 2013, pp. 1825--1830.

\bibitem{ollivier2019dsn}
S.~Ollivier, D.~Kline~Jr., R.~Kawsher, R.~Melhem, S.~Banja, and A.~K. Jones,
  ``Leveraging transverse reads to correct alignment faults in domain wall
  memories,'' in \emph{Proceedings of the IEEE/IFIP Dependable Systems and
  Networks Conference (DSN)}, Portland, OR, June 2019.

\bibitem{hifi}
C.~Zhang, G.~Sun, X.~Zhang, W.~Zhang, W.~Zhao, T.~Wang, Y.~Liang, Y.~Liu,
  Y.~Wang, and J.~Shu, ``Hi-fi playback: Tolerating position errors in shift
  operations of racetrack memory,'' in \emph{ACM SIGARCH Computer Architecture
  News}, vol. 43-3.\hskip 1em plus 0.5em minus 0.4em\relax ACM, 2015, pp.
  694--706.

\bibitem{archer2020foosball}
S.~Archer, G.~Mappouras, R.~Calderbank, and D.~Sorin, ``Foosball coding:
  Correcting shift errors and bit flip errors in 3d racetrack memory,'' in
  \emph{2020 50th Annual IEEE/IFIP International Conference on Dependable
  Systems and Networks (DSN)}.\hskip 1em plus 0.5em minus 0.4em\relax IEEE,
  2020, pp. 331--342.

\bibitem{kim2012case}
Y.~Kim, V.~Seshadri, D.~Lee, J.~Liu, and O.~Mutlu, ``A case for exploiting
  subarray-level parallelism (salp) in dram,'' in \emph{2012 39th Annual
  International Symposium on Computer Architecture (ISCA)}.\hskip 1em plus
  0.5em minus 0.4em\relax IEEE, 2012, pp. 368--379.

\bibitem{seshadri2013rowclone}
V.~Seshadri, Y.~Kim, C.~Fallin, D.~Lee, R.~Ausavarungnirun, G.~Pekhimenko,
  Y.~Luo, O.~Mutlu, P.~B. Gibbons, M.~A. Kozuch \emph{et~al.}, ``Rowclone: Fast
  and energy-efficient in-dram bulk data copy and initialization,'' in
  \emph{Proceedings of the 46th Annual IEEE/ACM International Symposium on
  Microarchitecture}, 2013, pp. 185--197.

\bibitem{yue2013exploiting}
J.~Yue and Y.~Zhu, ``Exploiting subarrays inside a bank to improve phase change
  memory performance,'' in \emph{2013 Design, Automation \& Test in Europe
  Conference \& Exhibition (DATE)}.\hskip 1em plus 0.5em minus 0.4em\relax
  IEEE, 2013, pp. 386--391.

\bibitem{TapeCache}
R.~Venkatesan, V.~Kozhikkottu, C.~Augustine, A.~Raychowdhury, K.~Roy, and
  A.~Raghunathan, ``Tapecache: a high density, energy efficient cache based on
  domain wall memory,'' in \emph{Proc. of ISLPED)}, 2012, pp. 185--190.

\bibitem{sun2013cross}
Z.~Sun, W.~Wu, and H.~Li, ``Cross-layer racetrack memory design for ultra high
  density and low power consumption,'' in \emph{2013 50th ACM/EDAC/IEEE Design
  Automation Conference (DAC)}.\hskip 1em plus 0.5em minus 0.4em\relax IEEE,
  2013, pp. 1--6.

\bibitem{LLG}
M.~R. Scheinfein and E.~A. Price, ``Llg user manual v2. 50,'' \emph{Code of the
  LLG simulator can be found at http://ligmicro. home. mindspring. com}, 2003.

\bibitem{lefevre-multiplication}
V.~Lefèvre, ``Multiplication by an integer constant,'' INRIA, Tech. Rep.
  RR-4192, 2001, ffinria-00072430f.

\bibitem{Bernstein-multiplication}
\BIBentryALTinterwordspacing
R.~Bernstein, ``Multiplication by integer constants,'' \emph{Software: Practice
  and Experience}, vol.~16, no.~7, pp. 641--652, 1986. [Online]. Available:
  \url{https://onlinelibrary.wiley.com/doi/abs/10.1002/spe.4380160704}
\BIBentrySTDinterwordspacing

\bibitem{pouchet2012polybench}
L.-N. Pouchet \emph{et~al.}, ``Polybench: The polyhedral benchmark suite,''
  \emph{URL: http://www. cs. ucla. edu/pouchet/software/polybench}, vol. 437,
  2012.

\bibitem{chan1998bitmap}
C.-Y. Chan and Y.~E. Ioannidis, ``Bitmap index design and evaluation,'' in
  \emph{Proceedings of the 1998 ACM SIGMOD international conference on
  Management of data}, 1998, pp. 355--366.

\bibitem{nvsim}
X.~Dong, C.~Xu, Y.~Xie, and N.~P. Jouppi, ``Nvsim: A circuit-level performance,
  energy, and area model for emerging nonvolatile memory,'' \emph{IEEE
  Transactions on Computer-Aided Design of Integrated Circuits and Systems},
  vol.~31, no.~7, pp. 994--1007, 2012.

\bibitem{nvsimRT}
C.~Zhang, G.~Sun, W.~Zhang, F.~Mi, H.~Li, and W.~Zhao, ``Quantitative modeling
  of racetrack memory, a tradeoff among area, performance, and power,'' in
  \emph{Design Automation Conference (ASP-DAC), 2015 20th Asia and South
  Pacific}.\hskip 1em plus 0.5em minus 0.4em\relax IEEE, 2015, pp. 100--105.

\bibitem{FreePDK45}
\BIBentryALTinterwordspacing
{North Carolina State University}, ``Freepdk45.'' [Online]. Available:
  \url{https://research.ece.ncsu.edu/eda/freepdk/freepdk45/}
\BIBentrySTDinterwordspacing

\bibitem{chen2010advances}
E.~Chen, D.~Apalkov, Z.~Diao, A.~Driskill-Smith, D.~Druist, D.~Lottis,
  V.~Nikitin, X.~Tang, S.~Watts, S.~Wang \emph{et~al.}, ``Advances and future
  prospects of spin-transfer torque random access memory,'' \emph{IEEE
  Transactions on Magnetics}, vol.~46, no.~6, pp. 1873--1878, 2010.

\bibitem{khan2019rtsim}
A.~A. Khan, F.~Hameed, R.~Bl{\"a}sing, S.~Parkin, and J.~Castrillon, ``Rtsim: A
  cycle-accurate simulator for racetrack memories,'' \emph{IEEE Computer
  Architecture Letters}, vol.~18, no.~1, pp. 43--46, 2019.

\bibitem{lenet}
Y.~LeCun, L.~Bottou, Y.~Bengio, and P.~Haffner, ``Gradient-based learning
  applied to document recognition,'' \emph{Proceedings of the IEEE}, vol.~86,
  no.~11, pp. 2278--2324, 1998.

\bibitem{alexnet}
A.~Krizhevsky, I.~Sutskever, and G.~E. Hinton, ``Imagenet classification with
  deep convolutional neural networks,'' \emph{Advances in neural information
  processing systems}, vol.~25, pp. 1097--1105, 2012.

\bibitem{jiang2019achieving}
W.~Jiang, E.~H.-M. Sha, X.~Zhang, L.~Yang, Q.~Zhuge, Y.~Shi, and J.~Hu,
  ``Achieving super-linear speedup across multi-fpga for real-time dnn
  inference,'' \emph{ACM Transactions on Embedded Computing Systems (TECS)},
  vol.~18, no.~5s, p.~67, 2019.

\bibitem{sim2018nid}
J.~Sim, H.~Seol, and L.-S. Kim, ``Nid: processing binary convolutional neural
  network in commodity dram,'' in \emph{2018 IEEE/ACM International Conference
  on Computer-Aided Design (ICCAD)}.\hskip 1em plus 0.5em minus 0.4em\relax
  IEEE, 2018, pp. 1--8.

\end{thebibliography}
\bibliographystyle{IEEEtran}


\end{document}